\documentclass[12]{article}
\usepackage{amsmath,amssymb}
\usepackage{epsfig}

\usepackage{color}

\def\Red#1{{\color{red}#1}}
\def\Green#1{{\color{darkgreen}#1}}
\def\Blue#1{{\color{blue}#1}}

\definecolor{red}{rgb}{1,0,0}
\definecolor{blue}{rgb}{0,0,1}
\definecolor{green}{rgb}{0,1,0}
\definecolor{black}{rgb}{0,0,0}
\definecolor{yellow}{rgb}{1,1,0}
\definecolor{mdwblue}{rgb}{0.2,0.2,0.6}
\definecolor{gray}{rgb}{0.7,0.7,0.7}
\definecolor{darkgreen}{rgb}{0.2,0.7,0.2}

\begin{document}
\author{Johan Noldus, Universiteit Gent \\* Vakgroep Wiskundige analyse, Galglaan 2, 9000 Gent, Belgium}

\title{The limit space of a Cauchy sequence of globally hyperbolic spacetimes}
\maketitle
\begin{abstract}
In this second paper, I construct a limit space of a Cauchy sequence of globally hyperbolic spacetimes.  In the second section, I work gradually towards a construction of the limit space.  I prove that the limit space is unique up to isometry.  
I also show that, in general, the limit space has quite complicated causal behaviour.  This work prepares the final paper in which I shall study in more detail properties of the limit space and the moduli space of (compact) globally hyperbolic spacetimes (cobordisms).  As a fait divers, I give in this paper a suitable definition of dimension of a Lorentz space in agreement with the one given by Gromov in the Riemannian case.       
\end{abstract}
\textsl{The difference in philosophy between Lorentzian and Riemannian geometry is one of relativism versus absolutism.  In the latter every point distinguishes itself while in the former in general two elements get distinguished by a third, different, one.}
\section{Introduction}
In this paper, I start from a slight modification of the Gromov-Hausdorff metric introduced in the previous one \cite{Noldus} and probe for a suitable construction of ``the'' limit space of a Cauchy sequence of spacetimes.  The modification (``convergence to invertibility'') seems necessary to me, since I was unable to prove the theorems in this paper without appealing to this property.  I would be grateful to the reader who could point out to me if ``convergence to invertibility'' is really needed or not.  As in the previous article, all spacetimes are compact (with spacelike boundaries) and globally hyperbolic.  Also, all conventions and notations are the same as in ~\cite{Noldus}.  
\newtheorem{deffie}{Definition}
\begin{deffie}
(\textbf{Modified Lorentzian Gromov-Hausdorff uniformity}) We call $(\mathcal{M},g)$ and $(\mathcal{N},h)$ $(\epsilon , \delta)$ close iff there exist mappings $\psi : \mathcal{M} \rightarrow \mathcal{N}$, $\zeta : \mathcal{N} \rightarrow \mathcal{M}$ such that 
\begin{eqnarray}
\left| d_{h} ( \psi ( p_{1} ), \psi (p_{2})) - d_{g} (p_{1} , p_{2} ) \right| & \leq & \epsilon \quad \forall p_{1}, p_{2} \in \mathcal{M} \\
\left| d_{g} ( \zeta ( q_{1} ) ,\zeta (q_{2})) - d_{h} ( q_{1} ,q_{2} )\right| & \leq & \epsilon \quad \forall q_{1}, q_{2} \in \mathcal{N} 
\end{eqnarray}
and
\begin{eqnarray} D_{\mathcal{M}}(p,\zeta \circ \psi(p)) & \leq & \delta  \\
D_{\mathcal{N}}(q, \psi \circ \zeta (q)) & \leq & \delta 
\end{eqnarray}
for all $p \in \mathcal{M}$ and $q \in \mathcal{N}$\footnote{$D_{\mathcal{M}}$ and $D_{\mathcal{N}}$ are the strong distances defined in \cite{Noldus}, definition $3$.  They are defined as $$ D_{\mathcal{M}} ( p , q ) = \max_{r \in \mathcal{M}} \left| d(p,r) + d(r,p) - d(q,r) - d(r,q) \right| $$}. $\square$
\end{deffie} 
\textbf{Remarks}: \\*
I show that the property $$ \left| d(p,r) + d(r,p) - d(q,r) - d(r,q) \right| < \epsilon $$ for all $r \in \mathcal{M}$ is equivalent to 
\begin{itemize}
\item $\left| d(p,r) - d(q,r) \right| < \epsilon$
\item $ \left| d(r,p) - d(r,q) \right| < \epsilon$
\end{itemize}  
for all $r \in \mathcal{M}$.   \\* \\*
\textsl{Proof}: \\*
$\Rightarrow )$ We prove only the first inequality, the second being analogous.  Without loss of generality we may assume that $d(p,r) > 0$.  Suppose $\epsilon > d(p,r) > 0$ and $d(q,r) \geq 0$ then $\left| d(p,r) - d(q,r) \right| < \epsilon $.  Suppose $d(p,r) \geq \epsilon$ then $d(q,r) > 0$ since if $d(r,q) >0$ then $d(p,q) > \epsilon$, hence 
$$ \left| d(p,q) + d(q,p) - 2d(q,q) \right| = d(p,q) > \epsilon $$
which is a contradiction.  But if $d(q,r) >0$, then again the first inequality follows. \\*
$\Leftarrow )$ Suppose that for some $r$
$$ \left| d(p,r) + d(r,p) - d(q,r) - d(r,q) \right| \geq \epsilon $$
Without loss of generality (all other cases are symmetric) we can assume that $d(p,r) > \epsilon$, hence $d(q,r) > 0$ otherwhise $\left| d(p,r) - d(q,r) \right| > \epsilon $ which is excluded by assumption.  But in this case
$$ \left| d(p,r) + d(r,p) - d(q,r) - d(r,q) \right| = \left| d(p,r) - d(q,r) \right| < \epsilon $$ which is a contradiction.  $\square$ \\* \\*
Using the above remark I show that Gromov-Hausdorff $(\epsilon, \delta)$ closeness has all the properties of a uniformity\footnote{An introduction to uniformities can be found in Appendix D.}.  Obviously, the notion is symmetric by definition and we are left to prove the generalized triangle inequality.  Hence, suppose that $(\mathcal{M}_{1} , g_{1})$ and $(\mathcal{M}_{2} , g_{2} )$ are $(\epsilon_{1}, \delta_{1} )$ close, assume also that $(\mathcal{M}_{2} , g_{2})$ and $(\mathcal{M}_{3} , g_{3} )$ are $(\epsilon_{2} , \delta_{2} )$ close, then $(\mathcal{M}_{1} , g_{1})$ and $(\mathcal{M}_{3} , g_{3} )$ are $( \epsilon_{1} + \epsilon_{2} , \delta_{1} + \delta_{2} + 2 \max \left\{ \epsilon_{1} , \epsilon_{2} \right\} )$ close.  \\* \\*
\textsl{Proof}:
Let $\psi_{i} : \mathcal{M}_{i} \rightarrow \mathcal{M}_{i+1}$ and $\zeta_{i} : \mathcal{M}_{i+1} \rightarrow \mathcal{M}_{i}$ be mappings which make $(\mathcal{M}_{i} , g_{i})$ and $(\mathcal{M}_{i+1} , g_{i+1} )$, $(\epsilon_{i}, \delta_{i} )$ close for $i=1,2$.  Then, we have for all $r,p \in \mathcal{M}_{1}$ that:
\begin{eqnarray*} \left| d_{g_{1}} ( \zeta_{1} \circ \zeta_{2} \circ \psi_{2} \circ \psi_{1} (p) , r ) - d_{g_{1}} ( p ,r ) \right| & \leq & \left| d_{g_{2}} ( \psi_{1} ( p ) , \psi_{1} (r)) - d_{g_{1}} ( p ,r ) \right| + \end{eqnarray*} 
\begin{eqnarray*} & & \left| d_{g_{1}} ( \zeta_{1} \circ \zeta_{2} \circ \psi_{2} \circ \psi_{1} (p) , r ) - d_{g_{2}} ( \psi_{1} \circ \zeta_{1} \circ \zeta_{2} \circ \psi_{2} \circ \psi_{1} ( p ) , \psi_{1} (r)) \right| +  \\ & &
\left| d_{g_{2}} ( \psi_{1} \circ \zeta_{1} \circ \zeta_{2} \circ \psi_{2} \circ \psi_{1} (p) , \psi_{1} (r) ) - d_{g_{2}} ( \zeta_{2} \circ \psi_{2} \circ \psi_{1} ( p ) , \psi_{1} (r)) \right| + \\ & &
\left| d_{g_{2}} ( \zeta_{2} \circ \psi_{2} \circ \psi_{1} ( p ) , \psi_{1} (r)) - d_{g_{2}} ( \psi_{1} ( p ) , \psi_{1} (r)) \right| 
\end{eqnarray*}     
Obviously, this implies that $$ \left| d_{g_{1}} ( \zeta_{1} \circ \zeta_{2} \circ \psi_{2} \circ \psi_{1} (p) , r ) - d_{g_{1}} ( p ,r ) \right|  \leq  2 \epsilon_{1} + \delta_{1} + \delta_{2}.$$  Making the same estimate for 
$$ \left| d_{g_{3}} ( \psi_{2} \circ \psi_{1} \circ \zeta_{1} \circ \zeta_{2} ( q ), s ) - d_{g_{3}} ( q ,s ) \right| $$ the result follows.  $\square$ \\* \\*
Let $(\mathcal{M}_{i} , g_{i})_{i = 1}^{n}$ be globally hyperbolic spacetimes such that $(\mathcal{M}_{i},g_{i})$ and $(\mathcal{M}_{i+1} , g_{i+1})$ are $(\epsilon_{i} , \delta_{i})$ close.  Then, in the same spirit as above, $(\mathcal{M}_{1} ,g_{1})$ and $(\mathcal{M}_{n} , g_{n} )$ are $(\sum_{i=1}^{n-1} \epsilon_{i}, \sum_{i=1}^{n-1} \delta_{i} + 2 \sum_{i = 2}^{n-2} \epsilon_{i} + 2 \max \left\{ \epsilon_{1} , \epsilon_{n-1} \right\})$ close.  As a consequence of theorem $6$ in ~\cite{Noldus}, $(\mathcal{M},g)$ and $(\mathcal{N},h)$ are isometric iff they cannot be distinguished by the the modified Gromov-Hausdorff uniformity.  Note also that it is impossible to make \emph{directly} a metric out of the $(\epsilon , \delta)$ Gromov-Hausdorff closeness, since if $\delta = f( \epsilon )$ where $f$ is a continuous function, one obtains for $\epsilon_{2} < \epsilon_{1}$ that
$$ f(\epsilon_{1}) + f(\epsilon_{2}) + 2 \epsilon_{1} \leq f(\epsilon_{1} + \epsilon_{2}) $$
which is impossible.              
\\*
\\*
The natural way to proceed now is to consider a generalized, Gromov-Hausdorff, Cauchy sequence of spacetimes, construct a ``reasonable'' limit space and finally deduce some properties of it.  This is the work done in the second and third section.  The presentation of this material is conservative, in the sense that, ab initio, the main goal is to construct a decent limit space using the Alexandrov topology.  This approach, however, turns out not to work since the candidate limit space is in general not $T_{2}$, the Lorentz distance $d$ is not continuous, nor is the limit space compact in this topology.  This part of the paper has its merits nevertheless, since the problems occuring bring alive important notions such as the timelike capacity, $\mathcal{TC}(\mathcal{M})$, and the timelike continuum $\mathcal{TCON}(\mathcal{M})$.  These notions express that in a limit space of a Cauchy sequence of globally hyperbolic cobordisms,  the definition of the causal relation from the chronological one is brought into jeopardy.  Hence, it is not clear if two chronologically related points can be connected by a causal curve (geodesic).  \\*
\\*
Since I want the limit $d$ to be continuous and the $T_{2}$ separation property to be valid (at least on the interior), the strong metric, which is already present in the definition of the modified, Gromov-Hausdorff uniformity itself, becomes important.  This metric turns out to have great technical potential, as shown in theorems 2,3 and 4.  Rafael Sorkin pointed out to me that David Meyer already had a similar idea, although for different purposes\cite{Meyer}.  In the last section, I study some examples which indicate what problems show up in an eventual definition of the causal relation.  It is also shown that for points $p \ll q$, the connecting geodesic (if it exists) is in general causal and not everywhere timelike.  Since the strong metric, $D_{\mathcal{M}}$, turned out to be such a strong device, I shall examine some first properties of it.  For example, it turns out that $D_{\mathcal{M}}$ cannot be a path metric.  Further study of the strong metric and causal curves is left for future work \cite{Noldus2}.  There is still a small philosophical remark which can be made about $D_{\mathcal{M}}$, it is a globally determined notion of locality.  This is clear from the definition itself, and the intermezzo at the end of section 3.\\* \\*                 
The obtained properties of the limit space give us a guideline for the abstract definition of a Lorentz space; this shall be done in a future paper \cite{Noldus2} and the line of thought displayed will be similar to the one followed by Busemann \cite{Busemann}.  In the epilogue I mention a good definition of dimension of an \emph{arbitrary} Lorentz space by comparing it with a Lorentzian manifold in the Gromov-Hausdorff sense.  \\*
\\*
As mentioned in the introduction of the previous paper, this work serves many physical fields such as cosmology and quantum gravity, since it establishes for the very first time an \emph{observer independent} way to compare spacetimes.  I admit that the abstract theory does not lend itself yet towards immediate application in these fields, but this will be considered in future work where the link with statistical Lorentzian geometry shall be examined.  
              
\section{Construction of the limit space}
Let $(\mathcal{M}_{i} , g_{i} )_{ i \in \mathbb{N}}$ be a sequence of compact, Lorentzian manifolds such that there exist mappings $\psi^{i}_{i+1} : \mathcal{M}_{i} \rightarrow \mathcal{M}_{i+1}$ and $\zeta^{i+1}_{i} : \mathcal{M}_{i+1} \rightarrow \mathcal{M}_{i}$ such that $\psi^{i}_{i+1}$ and $\zeta^{i+1}_{i}$ make $(\mathcal{M}_{i},g_{i})$ and $(\mathcal{M}_{i+1},g_{i+1})$, $(\frac{1}{2^{i}} ,\frac{1}{2^{i}} )$ close.  If we introduce the following mappings
$$ \psi^{i}_{i+k} = \psi^{i+k-1}_{i+k} \circ \psi^{i+k-2}_{i+k-1} \circ \ldots \circ \psi^{i+1}_{i+2} \circ \psi^{i}_{i+1} : \mathcal{M}_{i} \rightarrow \mathcal{M}_{i+k}$$   
$$ \zeta^{i+k}_{i} = \zeta^{i+1}_{i} \circ \zeta^{i+2}_{i+1} \circ \ldots \circ \zeta^{i+k-1}_{i+k-2} \circ \zeta^{i+k}_{i+k-1} : \mathcal{M}_{i+k} \rightarrow \mathcal{M}_{i}$$
then $\psi^{i}_{i+k}$ and $\zeta^{i+k}_{i}$ make $(\mathcal{M}_{i},g_{i})$ and $(\mathcal{M}_{i+k},g_{i+k})$, $(\frac{1}{2^{i-1}} , \frac{3}{2^{i-1}} )$ close.  Consider the set $\mathcal{S}$ of sequences $(x_{i})_{i \in \mathbb{N}}$, $x_{i} \in \mathcal{M}_{i
}$, such that there exists an $i_{0}$ such that for all $i > i_{0}$ one has that $x_{i} = \psi^{i_{0}}_{i} (x_{i_{0}})$.  Hence $x_{i} = \psi^{j}_{i} (x_{j})$ for all $i > j \geq i_{0}$.  
We make the following definition of Lorentzian distance:
$$ d((x_{i})_{i \in \mathbb{N}} , (y_{i})_{i \in \mathbb{N}} ) = \lim_{i \rightarrow \infty} d_{g_{i}} ( x_{i} , y_{i})$$
It is easy to verify that $d$ is well defined and a Lorentz distance.  The resulting partial order is defined as $(x_{i})_{i \in \mathbb{N}} \ll (y_{i})_{i \in \mathbb{N}}$ iff $d((x_{i})_{i \in \mathbb{N}} , (y_{i})_{i \in \mathbb{N}} )  > 0$.  Before we make a quotient of this space we should tell which topology is defined on it.  The obvious choice is the Alexandrov topology\footnote{We will see later on that this is a rather poor choice.} which is \textbf{generated} by the sets:
\begin{itemize}
\item $\mathcal{S}$, $\emptyset$
\item $I^{+} ((x_{i})_{i \in \mathbb{N}})$ and $I^{-} ((x_{i})_{i \in \mathbb{N}})$  
\item $I^{+} ((x_{i})_{i \in \mathbb{N}}) \cap I^{-} ((y_{i})_{i \in \mathbb{N}})$ with $d((x_{i})_{i \in \mathbb{N}} , (y_{i})_{i \in \mathbb{N}} ) > 0$ 
\end{itemize} 
with $I^{\pm}$ defined by the relation $\ll$. $\square$ \\*
\\*
\textbf{Remark} \\*
I stress the word ``generated'' since in general the above sets do \emph{not} constitute a basis of the Alexandrov topology as will become clear in the examples $2$ and $3$, where specific intersections of generating sets do not contain any generating set. $\square$  \\*
\\*  
As suggested in the previous article, in order to make sure that for any $(x_{i})_{i \in \mathbb{N}}$ there exists a point $(y_{i})_{i \in \mathbb{N}}$ such that $d((x_{i})_{i \in \mathbb{N}} , (y_{i})_{i \in \mathbb{N}} ) > 0$ or $d((y_{i})_{i \in \mathbb{N}} , (x_{i})_{i \in \mathbb{N}} ) > 0$ we have to demand that in the spaces $(\mathcal{M}_{i} ,g_{i})$ every point has a long enough past or a long enough future.  Therefore, we introduce the concept of timelike capacity.
\begin{deffie}
The timelike capacity $\mathcal{TC} (\mathcal{M},g)$ of a spacetime $(\mathcal{M} ,g)$ is defined as
$$ \mathcal{TC}(\mathcal{M},g) = \inf_{p \in \mathcal{M}} \sup_{q \in  \mathcal{M}} ( d_{g} ( p,q) + d_{g} (q,p) )$$
\end{deffie}   
Suppose now that the timelike capacity of the sequence $(\mathcal{M}_{i} ,g_{i})$ is bounded from below, i.e., there exists an $\alpha >0$ such that
$$ \mathcal{TC} ( \mathcal{M}_{i} ,g_{i} ) \geq \alpha \quad \forall i \in \mathbb{N}$$
then for any $(x_{i})_{i \in \mathbb{N}}$ one can find $(y_{i})_{i \in \mathbb{N}}$ such that the quantity $$d( (x_{i})_{i \in \mathbb{N}} , (y_{i})_{i \in \mathbb{N}} ) + d((y_{i})_{i \in \mathbb{N}} , (x_{i})_{i \in \mathbb{N}} )$$ can be chosen arbitrarily close to $\alpha$.\footnote{The reader for which this is not obvious is encouraged to make a formal proof.  This should go like: choose $\alpha > \epsilon > 0$ and $(x_{i})_{i \in \mathbb{N}} \in \mathcal{S}$. Let $\frac{1}{2^{i_{0}}} < \epsilon$, $i_{0} \in \mathbb{N}$ be such that $x_{j} = \psi_{j}^{i_{0}} (x_{i_{0}})$ for all $j > i_{0}$.  Let $y_{i_{0}} \in \mathcal{M}_{i_{0}}$ be such that $d_{g_{i_{0}}} (x_{i_{0}} , y_{i_{0}} ) + d_{g_{i_{0}}} ( y_{i_{0}} ,x_{i_{0}}) = \alpha$, then $\ldots$ }  If there is no control on the timelike capacity, this clearly needs not be case as is shown in the last example of the previous paper.  In what follows, I shall construct a candidate limit space and examine its separation properties in the Alexandrov topology.  It will turn out that the Alexandrov topology is too weak and the strong metric topology will emerge as a natural candidate.  \\* \\*
First, we show that one cannot expect any candidate limit space to be Hausdorff in the Alexandrov topology.  Suppose we allow that in the limit the boundary becomes a null surface, then obviously the limit space, equipped with the Alexandrov topology, is at most $T_{0}$.  This is illustrated by the next example.  \\*
\\*
\textbf{Example 1} \\*
Take the ``cylinder universe'' $S^{1} \times \mathbb{R}$ with metric $- dt^{2} + d\theta^{2}$ and let $p = (0,- \frac{T}{2})$ and $q = (0, \frac{T}{2})$ with $T > 0$. \\* In the notation of \cite{Beem} let $K^{+}(q,\epsilon) = \{ r | d(q,r) = \epsilon \}$ be the ``future ball'' of radius $\epsilon$ centred at $q$ and $K^{-}(p, \epsilon) = \{ r | d(r,p) = \epsilon \}$ be the ``past ball'' centred at $p$.  Consider the spacetimes $(J^{+} ( K^{-} (p, \epsilon)) \cap J^{-} ( K^{+} ( q, \epsilon )) , -dt^{2} + d\theta^{2} )$ then the \emph{unique} (up to isometry) Gromov-Hausdorff limit space for $\epsilon \rightarrow 0$ is $$(J^{+}(E^{-}(p)) \cap J^{-}( E^{+}(q)), -dt^{2} + d\theta^{2} )$$
which is $T_{0}$, but not $T_{1}$, in the Alexandrov topology.
\\*
\\*
However, this is, as will become clear later on, not only a \emph{boundary} phenomenon and in general the interior points of the $T_{0}$ quotient are not $T_{2}$ separated.  Let us first characterize the $T_{0}$ quotient of $\mathcal{S}$.
\newtheorem{theo}{Theorem}
\begin{theo}
The $T_{0}$ quotient $T_{0}\mathcal{S}$ of $\mathcal{S} = \{ (x_{i})_{i \in \mathbb{N}} \, | \, \exists i_{0} : \forall i \geq i_{0} \quad x_{i} = \psi_{i}^{i_{0}}( x_{i_{0}}) \}$ is defined by the equivalence relationship $\sim$ given by $(x_{i})_{i \in \mathbb{N}} \sim (y_{i})_{i \in \mathbb{N}}$ iff for all $(z_{i})_{i \in \mathbb{N}} \in \mathcal{S}$ 
\begin{equation} \label{ekkie} d( (x_{i})_{i \in \mathbb{N}} , (z_{i})_{i \in \mathbb{N}} ) + d((z_{i})_{i \in \mathbb{N}} , (x_{i})_{i \in \mathbb{N}} ) = d( (y_{i})_{i \in \mathbb{N}} , (z_{i})_{i \in \mathbb{N}} ) + d((z_{i})_{i \in \mathbb{N}} , (y_{i})_{i \in \mathbb{N}} ) \end{equation}  
\end{theo}  
\textsl{Proof}: \\*
Obviously, if two points $x,y \in \mathcal{S}$ are $T_{0}$ separated then there exists a $z \in \mathcal{S}$ such that (\ref{ekkie}) is not satisfied.  Suppose $x,y \in \mathcal{S}$ are not $T_{0}$ separated, but there exists a $z \in \mathcal{S}$ such that (\ref{ekkie}) is not satisfied.  Then there are essentialy two possibilities, either $d(z,x) > d(z,y)$ and $d(y,z) = 0$, or $d(x,z) > d(z,y)$ and $d(y,z) =0$.  The latter case implies that $x$ and $y$ are $T_{0}$ separated by $I^{-}(z)$ which is a contradiction (obviously $(\mathcal{S},d)$ is chronological).  The former case is proven by noticing that if $d(z,x) > d(z,y) + \delta$, then for $k$ sufficiently large such that $\frac{1}{2^{k-1}} < \frac{\delta}{8}$ and $z_{l} = \psi^{k}_{l} (z_{k})$ , $x_{l} =\psi^{k}_{l} (x_{k})$, $y_{l} = \psi^{k}_{l} (y_{k})$ for all $l > k$, one has that
$$ d_{g_{k}} (z_{k} , x_{k}) > d_{g_{k}} ( z_{k} , y_{k} ) + \frac{3 \delta}{4} $$ 
Choose $\gamma$ to be a distance maximizing geodesic in $\mathcal{M}_{k}$ from $z_{k}$ to $x_{k}$ and define the points $p_{k}$ and $q_{k}$ on $\gamma$ by
$$ d_{g_{k}}(p_{k} , q_{k}) = d_{g_{k}} ( q_{k} , x_{k} ) = \frac{3 \delta}{8} $$
then $p_{k}$ is not in the causal past of $y_{k}$.  Hence, $p = (p_{i})_{i \in \mathbb{N}}$ and $q = (q_{i})_{i \in \mathbb{N}}$, with $p_{i} = \psi_{i}^{k} (p_{k})$ and $q_{i} = \psi_{i}^{k} (q_{k})$ for all $i > k$, satisfy the properties \begin{itemize}
\item $d(p,y) < \frac{\delta}{8}$               
\item $d(p,q),d(q,x) > \frac{\delta}{4}$
\end{itemize}        
Hence, $q$ is not in the past of $y$, since otherwhise $d(p,y) \geq d(p,q) > \frac{\delta}{4}$ which is a contradiction.  But then, $x$ and $y$ are $T_{0}$ separated by $I^{+}(q)$ which is a contradiction.  $\square$
\\* 
\\*
Note that theorem 1 and the remark on page 2 imply that the Lorentzian distance $d$ is well defined on $T_{0}\mathcal{S}$ (that means independent of the representatives).  Theorem $1$ also reveals that the strong $T_{2}$ quotient of $\mathcal{S}$ equals $T_{0}\mathcal{S}$.  In the following few pages, I want to construct the \emph{timelike closure} of the $T_{0}$ quotient of $\mathcal{S}$.  Hence, I should first define timelike Cauchy sequences.
\begin{deffie}
A sequence $(x^{i})_{i \in \mathbb{N}}$ of points in $T_{0}\mathcal{S}$ is called future timelike Cauchy iff $x^{i} \ll x^{j}$ for all $i < j$ and $\forall \epsilon > 0, \, \exists i_{0}$ such that for all $k > j \geq i_{0}$ 
$$ 0 < d(x^{j} , x^{k}) < \epsilon $$
A past timelike Cauchy sequence is defined dually. $\square$
\end{deffie}   
Of course, some timelike Cauchy sequences determine the same ``limit point''.  Hence, I need to define when two timelike Cauchy sequences are \emph{equivalent}.
\begin{deffie}
Two future timelike Cauchy sequences $(x^{i})_{i \in \mathbb{N}}$, $(y^{i})_{i \in \mathbb{N}}$ in $T_{0}\mathcal{S}$ are equivalent iff for any $k$ there exists an $i_{0}$ such that $i \geq i_{0}$ implies that $x^{k} \ll y^{i}$ and $y^{k} \ll x^{i}$.  The equivalence relation for two past timelike Cauchy sequences is defined dually.  A future timelike Cauchy sequence $(x^{i})_{i \in \mathbb{N}}$ and a past timelike Cauchy sequence $(y^{i})_{i \in \mathbb{N}}$ are equivalent iff $x^{k} \ll y^{l}$ for any $k,l \in \mathbb{N}$ and there exist no two points $z^{1},z^{2} \in T_{0}\mathcal{S}$ such that 
$$ x^{k} \ll z^{1} \ll z^{2} \quad \forall k \textrm{ and } z^{2} \notin \bigcup_{j \in \mathbb{N}} I^{+} ( y^{j}) $$
or
$$ y^{k} \gg z^{1} \gg z^{2} \quad \forall k \textrm{ and } z^{2} \notin \bigcup_{j \in \mathbb{N}} I^{-} (x^{j}) $$
$\square$ 
\end{deffie}
\textbf{Genesis}: \\*
The only point in the above definition which might not be obvious for some readers is why there are two points $z^{1},z^{2}$ included in the definition of equivalence between a future $(p^{i})_{i \in \mathbb{N}}$ and past timelike $(q^{i})_{i \in \mathbb{N}}$ Cauchy sequence.  Consider a spacetime which is already timelike complete (such as Minkowski spacetime), and let $z^{1}$ be the limit point of the sequence $(p^{i})_{i \in \mathbb{N}}$.  Then, the only conclusion which one can draw from $p^{i} \ll q^{j}$ for all $i,j > 0$ is that $z^{1}$ is null connected to the limit point of the sequence $(q^{i})_{i \in \mathbb{N}}$ which implies that $z^{1} \notin \bigcup_{j \in \mathbb{N}} I^{+} ( q^{j})$ but still the limit point of $(q^{i})_{i \in \mathbb{N}}$ might coincide with the limit point of $(p^{i})_{i \in \mathbb{N}}$.  Clearly, if there would exist a second point $z^{2}$ satisfying $z^{1} \ll z^{2} \notin \bigcup_{j \in \mathbb{N}} I^{+} ( q^{j})$, then the limit points cannot coincide.  $\square$  \\*
\\*  
It is left as an exercise to the reader that this definition determines an equivalence relation (see Appendix A).  I construct now the timelike closure, $\overline{T_{0}\mathcal{S}}$, of $T_{0}\mathcal{S}$.  Define $\widetilde{T_{0}\mathcal{S}}$ as the union of $T_{0}\mathcal{S}$ with all timelike Cauchy sequences in $T_{0}\mathcal{S}$.  Topologize $\widetilde{T_{0}\mathcal{S}}$ as follows.  $\mathcal{O} \subset \widetilde{T_{0}\mathcal{S}}$ is a \textbf{generating} set iff
\begin{itemize}
\item $\mathcal{O} \cap T_{0}\mathcal{S}$ is a generating set for the Alexandrov topology in $T_{0}\mathcal{S}$
\item A future (past) timelike Cauchy sequence $(p^{i})_{i \in \mathbb{N}}$ in $T_{0}\mathcal{S}$ belongs to $\mathcal{O}$ iff 
\begin{itemize}
\item $\mathcal{O} \cap T_{0}\mathcal{S} = I^{+} (q)$ ($I^{-}(q)$) for some $q \in T_{0}\mathcal{S}$ and there exists an $i_{0} \in \mathbb{N}_{0}$ such that $i \geq i_{0}$ implies that $p^{i} \in \mathcal{O} \cap T_{0}\mathcal{S} $, or
\item there exist $r^{1} , r^{2} \in T_{0}\mathcal{S} \cap \mathcal{O}$, $i_{0} \in \mathbb{N}_{0}$ such that $p^{i} \ll r^{2} \ll r^{1}$  ($r^{1} \ll r^{2} \ll p^{i}$) and $p^{i} \in \mathcal{O} \cap T_{0}\mathcal{S}$ for all $i \geq i_{0}$. 
\end{itemize}
\end{itemize}
$\overline{T_{0}\mathcal{S}}$ is defined as the $T_{0}$ quotient (in the Alexandrov topology) of $\widetilde{T_{0}\mathcal{S}}$\footnote{It will become clear later on that the definition of $\overline{T_{0}\mathcal{S}}$ is dependent on the mappings $\psi$ and $\zeta$ used to construct $\mathcal{S}$.}.
\\*
\\*
\textbf{Remark}:
\\*
Again, the intersection of two generating sets is in general \emph{not} equal to some union of generating sets as example $2$ shows.  After having studied examples $2$ and $3$, the reader should get a taste for the reason why the above definition is constructed in such a delicate way.  $\square$  \\*
\\*  
\textbf{Property}:
Two timelike Cauchy sequences are $T_{0}$ separated iff they are inequivalent.  \\*
$\Leftarrow )$ Suppose $(p^{i})_{i \in \mathbb{N}}$ and $(q^{i})_{i \in \mathbb{N}}$ are future timelike inequivalent.  Then, there exists a $k$ and a sequence $(l_{n})_{n \in \mathbb{N}}$ such that, say, $p^{k}$ is not in the timelike past of $q^{l_{n}}$ for any $n \in \mathbb{N}$.  But then $p^{k}$ is not in the past of any $q^{i}$ with $i \geq l_{0}$, which implies that $I^{+} (p^{k})$ contains $(p^{i})_{i \in \mathbb{N}}$ but not $(q^{i})_{i \in \mathbb{N}}$.  Suppose now that $(p^{i})_{i \in \mathbb{N}}$ is future timelike Cauchy and $(q^{i})_{i \in \mathbb{N}}$ is past timelike Cauchy, with $(p^{i})_{i \in \mathbb{N}}$ not equivalent to $(q^{i})_{i \in \mathbb{N}}$.  There are essentially two cases:
\begin{itemize}
\item there exist $k,l$ such that $p^{k}$ is not in the timelike past of $q^{l}$, but then $p^{k}$ is not in the timelike past of all $q^{s}$ for all $s \geq l$.  Hence $I^{+} (p^{k})$ separates $(p^{i})_{i \in \mathbb{N}}$ from $(q^{i})_{i \in \mathbb{N}}$.
\item $p^{k} \ll q^{l}$ for all $k,l \in \mathbb{N}$ but there exist $z^{1}, z^{2} \in \mathcal{M}$ such that, say, $ p^{k} \ll z^{1} \ll z^{2}$ but $z^{2} \notin \bigcup_{l \in \mathbb{N}} I^{+} (q^{l})$; then clearly $I^{-}(z^{2})$ separates $(p^{i})_{i \in \mathbb{N}}$ from $(q^{i})_{i \in \mathbb{N}}$.
\end{itemize}
$\Rightarrow )$ Suppose $(p^{i})_{i \in \mathbb{N}}$ and $(q^{i})_{i \in \mathbb{N}}$ are future timelike equivalent, then every set of the form $I^{+}(r)$ contains an element $p^{i}$ iff it contains an element $q^{j}$ and hence all $p^{s}$ for all $s \geq i$ and $q^{t}$ for all $t \geq j$, and the same property is valid for a finite number of intersections of such sets.  A set of the form $I^{-}(r)$ contains $(p^{i})_{i \in \mathbb{N}}$ iff there exists a $r^{1}$ with $p^{i} \ll r^{1} \ll r$ for all $i \in \mathbb{N}$.  Hence, $q^{i} \ll r^{1}$ for all $i \in \mathbb{N}$ and therefore $(q^{i})_{i \in \mathbb{N}} \in I^{-}(r)$.  Hence, $(p^{i})_{i \in \mathbb{N}}$ and $(q^{i})_{i \in \mathbb{N}}$ are $T_{0}$ equivalent.  Suppose now that $(p^{i})_{i \in \mathbb{N}}$ is future timelike Cauchy and $(q^{i})_{i \in \mathbb{N}}$ is past timelike Cauchy, with $(p^{i})_{i \in \mathbb{N}}$ equivalent to $(q^{i})_{i \in \mathbb{N}}$.  Let $I^{-}(r)$ be an open set containing $(p^{i})_{i \in \mathbb{N}}$.  Then, there exists an $r^{1}$ such that $p^{i} \ll r^{1} \ll r$ for all $i \in \mathbb{N}$.  Consequently, there exists a $j_{0}$ such that $j \geq j_{0}$ implies that $q^{j}$ belongs to $I^{-}(r)$, hence $(q^{j})_{j \in \mathbb{N}} \in I^{-}(r)$.  Hence, every Alexandrov set containing $(p^{i})_{i \in \mathbb{N}}$ contains $(q^{j})_{j \in \mathbb{N}}$.  The symmetric case is proven identically.         
$\square$     \\*    
\\*
Define the \emph{timelike continuum} of $\overline{T_{0}\mathcal{S}}$ as the subset of all points $r$, such there exist timelike past \emph{and} future Cauchy sequences in $T_{0}\mathcal{S}$ which are $T_{0}$ equivalent with $r$ in the Alexandrov topology.  \\* \\*
\textbf{Remark}: \\*
It is not true that if future timelike $(p^{i})_{i \in \mathbb{N}}$ and past timelike Cauchy sequences $(q^{i})_{i \in \mathbb{N}}$ converge to $r$ in the Alexandrov topology, then $r$, $(p^{i})_{i \in \mathbb{N}}$ and $(q^{i})_{i \in \mathbb{N}}$ are $T_{0}$ equivalent.  Moreover, $(p^{i})_{i \in \mathbb{N}} \sim (q^{i})_{i \in \mathbb{N}}$ does not imply that $r$ is $T_{0}$ equivalent with $(p^{i})_{i \in \mathbb{N}} \sim (q^{i})_{i \in \mathbb{N}}$.  This is illustrated in example $4$. $\square$ \\*
\\*
The next examples show that the Alexandrov topology is too weak.
\\*    
Moreover, in general the timelike continuum is a proper subset of $\overline{T_{0}\mathcal{S}}$.  I will also give an example in which every maximal $T_{2}$ subspace coincides, apart from a few points, with the timelike continuum.   \\* \\*
\textbf{Example 2}  \\*
This example is meant as a technical warm-up, and shows the mildest form of ``exotic'' behaviour.  We study on $S^{1} \times \left[ 0 , 5 \right]$ a family of conformally equivalent\footnote{Two metrics $d_{1}$ and $d_{2}$ on the same underlying space $\mathcal{N}$ are conformally equivalent iff $d_{1}(x,y) > 0 \Leftrightarrow d_{2} (x,y) > 0$ for all $x,y \in \mathcal{N}$.} Lorentz metrics $d_{\epsilon}$, whose limit distance $d$ is ``entirely degenerate'' on the strip $\mathcal{U} = S^{1} \times \left[ 1 , 1 + \pi \right]$, i.e., $d(x,y) = 0$ for all $x,y \in \mathcal{U}$.  However, it turns out that all points of $\mathcal{U}$ \emph{are} $T_{2}$ separated. This is a consequence of the ``external'' relations with points outside $\mathcal{U}$ and because the strip $\mathcal{U}$ is not ``thick enough'' (this will be made clear later on).  This stands diametrically opposite to what we are used to from pseudo\footnote{Pseudo in the sense that the ``Riemannian'' metric is allowed to be degenerate.} Riemannian geometry in which this is just impossible because of the triangle inequality \footnote{Let $\mathcal{U}$ be a maximal set in a pseudo ``Riemannian'' space on which the ``Riemannian'' metric $D$ vanishes, then for any exterior point $r$ and $x,y \in \mathcal{U}$ such that $D(r,x)$ or $D(r,y)$ is nonvanishing, one has that $D(r,x) \leq D(r,y) \leq D(r,x)$.}.  In \emph{global} Lorentzian geometry however, the reversed triangle inequality is responsible for this phenomenon.  I am aware that some classical relativists might start objecting to the construction now, but let me convince these people that it is their own, highly Riemannesque, ``local intuition'' which is responsible for this protest.  Moreover this global ``artifact'' has as peculiarity that we are still able to reconstruct the ``faithful'' causal relations even on the strip $\mathcal{U}$ of degeneracy.  In what follows, I construct conformal factors $\Omega_{\epsilon}(t)$ which equal $1$ on $\left[0 , 1 - \epsilon \right] \cup \left[ 1 + \pi + \epsilon , 5 \right]$, $\epsilon$ on $\left[ 1 + \epsilon , 1 + \pi - \epsilon \right]$ and undergo a smooth transition in the strips $\left[ 1 - \epsilon , 1 + \epsilon \right]$ and $\left[ 1 + \pi - \epsilon , 1 + \pi + \epsilon \right]$ respectively.  Such a smooth transition function can be constructed by using the smooth function $f$, defined by $f(x) = 0$ for $x \leq 0$ and $f(x) = \exp^{ - \frac{1}{x^{2}}}$ for $ x > 0$.  Define moreover $\alpha_{\epsilon} (x) = 1 - (1 - \epsilon) \chi (x - \epsilon)$, $\beta_{\epsilon}(x) = \epsilon + ( 1 - \epsilon ) \chi ( x - \epsilon )$ where $\chi$ is the characteristic function defined by $\chi (x ) = 1$ if $x > 0$ and zero otherwise.  Then with $$\psi_{\epsilon} ( x ) = \frac{ \alpha_{\epsilon}(\cdot - 1 + \epsilon ) \ast f(\cdot + \epsilon)f( - \cdot + \epsilon  ) }{ \int_{- \infty}^{+\infty} f(t + \epsilon) f(-t + \epsilon) dt} ( x)$$ and
$$ \zeta_{\epsilon} ( x ) = \frac{\beta_{\epsilon}(\cdot - 1 -\pi + \epsilon) \ast f( \cdot + \epsilon  )f( - \cdot + \epsilon )   }{ \int_{-\infty}^{+\infty} f(t + \epsilon)f( - t + \epsilon) dt }(x)$$        
one can define for example that $\Omega_{\epsilon}(t) = \psi_{\epsilon} (t) $ for $t \in \left[ 1 - \epsilon , 1 + \epsilon \right]$ and $\Omega_{\epsilon}(t) = \zeta_{\epsilon} (t)$ on $\left[ 1 + \pi - \epsilon , 1 + \pi + \epsilon \right]$. (In the above formula's, $\ast$ denotes the convolution product.)
\begin{figure}
\begin{center}
\mbox{\rotatebox{-90}{\includegraphics[width=4cm,height=6cm]{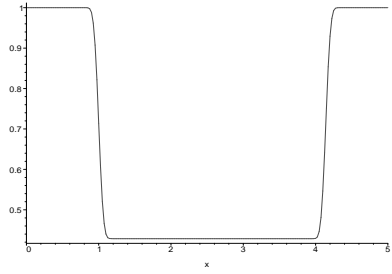}}}
\caption{A plot of $\Omega$. }
\label{fig1}
\end{center}
\end{figure}
Consider now the metric tensors
$$ ds^{2}_{\epsilon} = \Omega_{\epsilon}^{2}(t) ( -dt^{2} + d\theta^{2} )$$ and the associated Lorentz distances $d_{\epsilon}$.   On basis of simple geometric arguments, (i.e., without even calculating the geodesics) it is easy to see that $(S^{1} \times \left[0,5 \right], d_{\epsilon})$ is a Gromov-Hausdorff Cauchy sequence (with the maps $\psi_{\delta}^{\epsilon}$ and $\zeta_{\epsilon}^{\delta}$ equal to the identity) converging to the cylindrical space with a Lorentz metric $d$ which is degenerate on the strip $\left[ 1 , 1 + \pi \right]$ (Hint: look for lower and upper bounds of lengths of geodesics).  However, all the points on the strip are $T_{2}$ separated as is shown (partially) in Fig.\ \ref{fig2}.
\begin{figure} 
\begin{center}
  \setlength{\unitlength}{1.8cm}
\begin{picture}(8,5)

\put(1,1){\line(1,0){6}}
\put(1,1){\line(0,1){3}}
\put(1,4){\line(1,0){6}}
\put(7,1){\line(0,1){3}}
\put(1,0.8){$0$}
\put(7,0.8){$2\pi$}
\put(0.8,4){$\pi$}

\thicklines
\put(1,3){\line(1,1){1.3}}
\put(7,3){\line(-1,-1){1.7}}
\put(5.3,1.3){\line(-1,1){3}}
\put(2.3,4.4){$q_1$}

\put(2.3,1.3){\Red{\line(1,1){3}}}
\put(2.3,1.3){\Red{\line(-1,1){1.3}}}
\put(5.3,4.3){\Red{\line(1,-1){1.7}}}
\put(5.3,4.4){$q_2$}

\put(1.2,3.2){\Green{\line(1,-1){2.4}}}
\put(3.6,0.8){\Green{\line(1,1){1.1}}}
\put(4.2,3.2){\Green{\line(1,-1){2.4}}}
\put(6.6,0.8){\Green{\line(1,1){0.4}}}
\put(1,1.2){\Green{\line(1,1){0.7}}}
\put(3.6,0.6){$p_1$}
\put(6.6,0.6){$p_2$}

\thinlines
\multiput(1,3)(0.25,0){24}{\line(1,0){0.125}}
\multiput(1,2.6)(0.25,0){24}{\Red{\line(1,0){0.125}}}
\multiput(1,1.2)(0.25,0){24}{\Green{\line(1,0){0.125}}}

\multiput(1.3,3.1)(0.11,-0.11){21}{\Blue{\line(1,1){1.1}}}
\multiput(4.3,3.1)(0.11,-0.11){15}{\Blue{\line(1,1){1.1}}}
\put(5.95, 1.45){\Blue{\line(1,1){1.05}}}
\put(6.06, 1.34){\Blue{\line(1,1){0.94}}}
\put(6.17, 1.23){\Blue{\line(1,1){0.83}}}
\put(1, 2.06){\Blue{\line(1,1){0.27}}}
\put(6.28, 1.12){\Blue{\line(1,1){0.72}}}
\put(1, 1.84){\Blue{\line(1,1){0.38}}}
\put(6.39, 1.01){\Blue{\line(1,1){0.61}}}
\put(1, 1.62){\Blue{\line(1,1){0.49}}}
\put(6.50, 0.9){\Blue{\line(1,1){0.50}}}
\put(1, 1.4){\Blue{\line(1,1){0.60}}}

\put(2.5,2.3){\circle*{0.05}}
\put(2.6,2.3){$r_1$}
\put(5.5,2.3){\circle*{0.05}}
\put(5.6,2.3){$r_2$}

\end{picture}
\caption{Points $r_{1},r_{2}$ in the degenerate area can be Hausdorff separated.}
\label{fig2}
\end{center}
\end{figure}
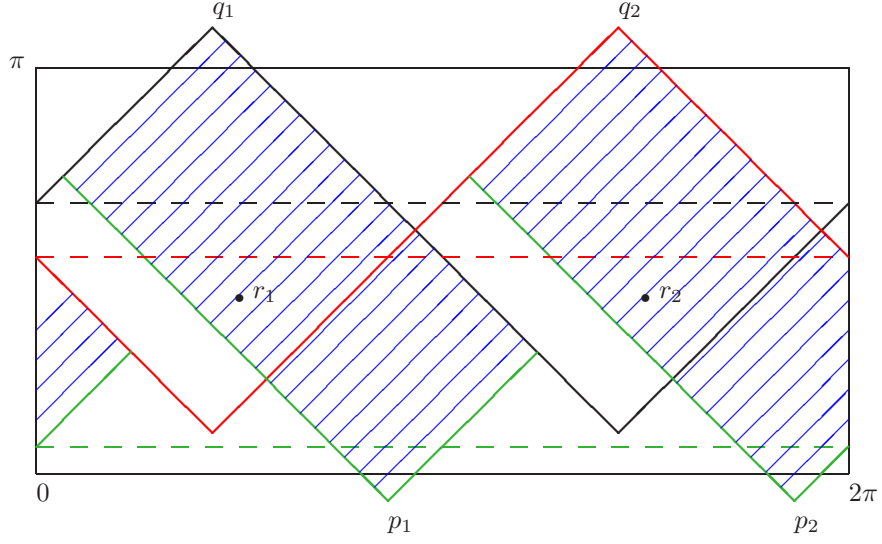
This will not be the case anymore in the next example.  The reader should do the following exercises in order to get used to the ``strange'' things which can happen. 
\begin{itemize}
\item Construct an open set in the ``degenerate area'' which does not contain any set of the form $I^{+}(p) \cap I^{-} (q)$ for $p \ll q$, elements of $\overline{T_{0}\mathcal{S}}$.
\item Consider the past timelike Cauchy sequence $((0, 1 + \pi + \frac{1}{n+1}))_{n \in \mathbb{N}}$, construct two \emph{generating} Alexandrov sets (in $\overline{T_{0}\mathcal{S}}$) $I^{+}(p_{i}) \cap I^{-}(q_{i})$, $i=1,2$ such that there exist $z_{i}$ with $p_{i} \ll z_{i} \ll (0,1 + \pi + \frac{1}{n+1})$ for all 
$n \in \mathbb{N}$ \emph{but} there exists no point $z \in I^{+}(p_{1}) \cap I^{+}(p_{2}) \cap I^{-}(q_{1}) \cap I^{-}(q_{2})$ such that $z \ll (0, 1 + \pi + \frac{1}{n+1})$ for all $n \in \mathbb{N}$.  
\end{itemize}
 $\square$
\\*
\\*
\textbf{Example 3}
This more exotic example shall make all points in the ``degenerate area'' only $T_{0}$ and \emph{not} $T_{2}$ separated.
 Consider, again, the cylinder $S^{1} \times \left[ 0 ,8 \right]$ and define conformal factors $\Omega_{\epsilon} (t)$ as follows: $\Omega_{\epsilon}(t)$ equals $1$ on the strips $\left[ 0 ,1 - \epsilon \right]$ and $\left[ 1 + 2 \pi + \epsilon , 8 \right]$, $\epsilon$ on $\left[ 1 + \epsilon , 1 + 2 \pi - \epsilon \right]$, and undergoes smooth transitions on $\left[ 1 - \epsilon , 1 + \epsilon \right]$ and $\left[ 1 + 2 \pi - \epsilon , 1 + 2 \pi + \epsilon \right]$ respectively.  Again, it is not difficult to show that the distances $d_{ \epsilon}$ associated with the metrics
$$ ds^{2}_{\epsilon} = \Omega_{\epsilon}^{2}(t) ( -dt^{2} + d\theta^{2}) $$  
form a Gromov-Hausdorff Cauchy sequence, and define a limit distance $d$ (wrt. to the identity mappings $\psi^{\epsilon}_{\delta}$ and $\zeta
^{\delta}_{\epsilon}$ ) which is degenerate on the strip $\left[1 , 1 + 2 \pi \right]$\footnote{$(S^{1} \times \left[ 0 ,8 \right] , d)$ is timelike complete.}.  However, the points with time coordinate $t = 1 + \pi$ are $T_{0}$ equivalent and therefore the $T_{0}$ limit space is topologically a (double) cone with (common) tip $t = 1 + \pi$.  Points in the strip $\left[ 1 , 1 + 2 \pi \right] $ cannot be $T_{2}$ separated, since any Alexandrov set containing such point also contains the tip\footnote{The $T_{1}$ separation property is also not satisfied since no point of the degenerate strip can be $T_{1}$ separated from the tip.} (see Fig.\ \ref{fig3}).
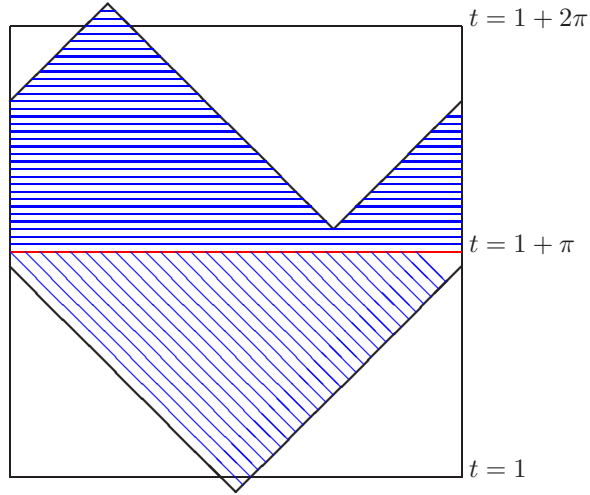
\begin{figure}[h]
\begin{center}
  \setlength{\unitlength}{1cm}
\begin{picture}(8,7)

\put(1,1){\line(1,0){6}}
\put(1,1){\line(0,1){6}}
\put(1,7){\line(1,0){6}}
\put(7,1){\line(0,1){6}}
\put(7.1,1){$t=1$}
\put(7.1,4){$t=1+\pi$}
\put(7.1,7){$t=1+2\pi$}
\put(1,4){\Red{\line(1,0){6}}}

\thicklines
\put(1,6){\line(1,1){1.3}}
\put(7,6){\line(-1,-1){1.7}}
\put(5.3,4.3){\line(-1,1){3}}

\put(1,3.8){\line(1,-1){3}}
\put(4,0.8){\line(1,1){3}}

\thinlines
\put(1,4){\Blue{\line(1,-1){3.1}}}
\put(1.2,4){\Blue{\line(1,-1){3}}}
\put(1.4,4){\Blue{\line(1,-1){2.9}}}
\put(1.6,4){\Blue{\line(1,-1){2.8}}}
\put(1.8,4){\Blue{\line(1,-1){2.7}}}
\put(2,4){\Blue{\line(1,-1){2.6}}}
\put(2.2,4){\Blue{\line(1,-1){2.5}}}
\put(2.4,4){\Blue{\line(1,-1){2.4}}}
\put(2.6,4){\Blue{\line(1,-1){2.3}}}
\put(2.8,4){\Blue{\line(1,-1){2.2}}}
\put(3,4){\Blue{\line(1,-1){2.1}}}
\put(3.2,4){\Blue{\line(1,-1){2}}}
\put(3.4,4){\Blue{\line(1,-1){1.9}}}
\put(3.6,4){\Blue{\line(1,-1){1.8}}}
\put(3.8,4){\Blue{\line(1,-1){1.7}}}
\put(4,4){\Blue{\line(1,-1){1.6}}}
\put(4.2,4){\Blue{\line(1,-1){1.5}}}
\put(4.4,4){\Blue{\line(1,-1){1.4}}}
\put(4.6,4){\Blue{\line(1,-1){1.3}}}
\put(4.8,4){\Blue{\line(1,-1){1.2}}}
\put(5,4){\Blue{\line(1,-1){1.1}}}
\put(5.2,4){\Blue{\line(1,-1){1}}}
\put(5.4,4){\Blue{\line(1,-1){0.9}}}
\put(5.6,4){\Blue{\line(1,-1){0.8}}}
\put(5.8,4){\Blue{\line(1,-1){0.7}}}
\put(6,4){\Blue{\line(1,-1){0.6}}}
\put(6.2,4){\Blue{\line(1,-1){0.5}}}
\put(6.4,4){\Blue{\line(1,-1){0.4}}}
\put(6.6,4){\Blue{\line(1,-1){0.3}}}
\put(6.8,4){\Blue{\line(1,-1){0.2}}}
\put(1,4.1){\Blue{\line(1,0){6}}}
\put(1,4.2){\Blue{\line(1,0){6}}}
\put(1,4.3){\Blue{\line(1,0){6}}}
\put(1,4.4){\Blue{\line(1,0){4.2}}}
\put(1,4.5){\Blue{\line(1,0){4.1}}}
\put(1,4.6){\Blue{\line(1,0){4}}}
\put(1,4.7){\Blue{\line(1,0){3.9}}}
\put(1,4.8){\Blue{\line(1,0){3.8}}}
\put(1,4.9){\Blue{\line(1,0){3.7}}}
\put(1,5){\Blue{\line(1,0){3.6}}}
\put(1,5.1){\Blue{\line(1,0){3.5}}}
\put(1,5.2){\Blue{\line(1,0){3.4}}}
\put(1,5.3){\Blue{\line(1,0){3.3}}}
\put(1,5.4){\Blue{\line(1,0){3.2}}}
\put(1,5.5){\Blue{\line(1,0){3.1}}}
\put(1,5.6){\Blue{\line(1,0){3}}}
\put(1,5.7){\Blue{\line(1,0){2.9}}}
\put(1,5.8){\Blue{\line(1,0){2.8}}}
\put(1,5.9){\Blue{\line(1,0){2.7}}}
\put(1,6){\Blue{\line(1,0){2.6}}}
\put(1.1,6.1){\Blue{\line(1,0){2.4}}}
\put(1.2,6.2){\Blue{\line(1,0){2.2}}}
\put(1.3,6.3){\Blue{\line(1,0){2}}}
\put(1.4,6.4){\Blue{\line(1,0){1.8}}}
\put(1.5,6.5){\Blue{\line(1,0){1.6}}}
\put(1.6,6.6){\Blue{\line(1,0){1.4}}}
\put(1.7,6.7){\Blue{\line(1,0){1.2}}}
\put(1.8,6.8){\Blue{\line(1,0){1}}}
\put(1.9,6.9){\Blue{\line(1,0){0.8}}}
\put(2,7){\Blue{\line(1,0){0.6}}}
\put(2.1,7.1){\Blue{\line(1,0){0.4}}}
\put(2.2,7.2){\Blue{\line(1,0){0.2}}}
\put(7,4.4){\Blue{\line(-1,0){1.6}}}
\put(7,4.5){\Blue{\line(-1,0){1.5}}}
\put(7,4.6){\Blue{\line(-1,0){1.4}}}
\put(7,4.7){\Blue{\line(-1,0){1.3}}}
\put(7,4.8){\Blue{\line(-1,0){1.2}}}
\put(7,4.9){\Blue{\line(-1,0){1.1}}}
\put(7,5){\Blue{\line(-1,0){1}}}
\put(7,5.1){\Blue{\line(-1,0){0.9}}}
\put(7,5.2){\Blue{\line(-1,0){0.8}}}
\put(7,5.3){\Blue{\line(-1,0){0.7}}}
\put(7,5.4){\Blue{\line(-1,0){0.6}}}
\put(7,5.5){\Blue{\line(-1,0){0.5}}}
\put(7,5.6){\Blue{\line(-1,0){0.4}}}
\put(7,5.7){\Blue{\line(-1,0){0.3}}}
\put(7,5.8){\Blue{\line(-1,0){0.2}}}

\end{picture}
\caption{Picture of the candidate limit space.}
\label{fig3}
\end{center}
\end{figure}

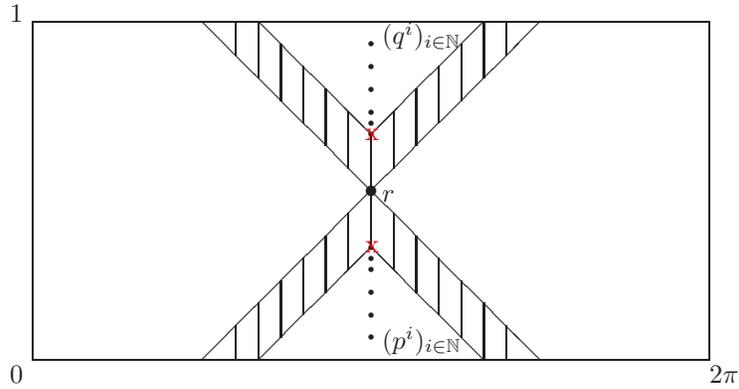
\begin{figure}[h]
\begin{center}
  \setlength{\unitlength}{1.5cm}
\begin{picture}(8,5)

\put(1,1){\line(1,0){6}}
\put(1,1){\line(0,1){3}}
\put(1,4){\line(1,0){6}}
\put(7,1){\line(0,1){3}}
\put(0.8,0.8){$0$}
\put(7,0.8){$2\pi$}
\put(0.8,4){$1$}

\put(2.5,1){\line(1,1){3}}
\put(2.5,4){\line(1,-1){3}}
\put(4,2.5){\circle*{0.1}}
\put(4.1,2.4){$r$}
\put(3,1){\line(1,1){1}}
\put(5,1){\line(-1,1){1}}
\put(3,4){\line(1,-1){1}}
\put(5,4){\line(-1,-1){1}}

\multiput(0,0)(0.2,0.2){6}{\put(3,1){\line(0,1){0.5}}}
\put(2.8,1){\line(0,1){0.3}}
\multiput(0,0)(-0.2,0.2){5}{\put(5,1){\line(0,1){0.5}}}
\put(5.2,1){\line(0,1){0.3}}
\multiput(0,0)(0.2,-0.2){6}{\put(3,4){\line(0,-1){0.5}}}
\put(2.8,4){\line(0,-1){0.3}}
\multiput(0,0)(-0.2,-0.2){5}{\put(5,4){\line(0,-1){0.5}}}
\put(5.2,4){\line(0,-1){0.3}}

\put(4.1,3.8){$(q^i)_{i\in\mathbb{N}}$}
\put(4,3.8){\circle*{0.05}}
\put(4,3.6){\circle*{0.05}}
\put(4,3.4){\circle*{0.05}}
\put(4,3.2){\circle*{0.05}}
\put(4,3.1){\circle*{0.05}}
\put(4,3.0){\circle*{0.05}}
\put(3.95,2.95){\Red{x}}
\put(4.1,1.1){$(p^i)_{i\in\mathbb{N}}$}
\put(4,1.2){\circle*{0.05}}
\put(4,1.4){\circle*{0.05}}
\put(4,1.6){\circle*{0.05}}
\put(4,1.8){\circle*{0.05}}
\put(4,1.9){\circle*{0.05}}
\put(4,2){\circle*{0.05}}
\put(3.95,1.95){\Red{x}}

\end{picture}
\caption{Convergence in the Alexandrov topology is not the same as $T_{0}$ equivalence. }
\label{fig4}
\end{center}
\end{figure}

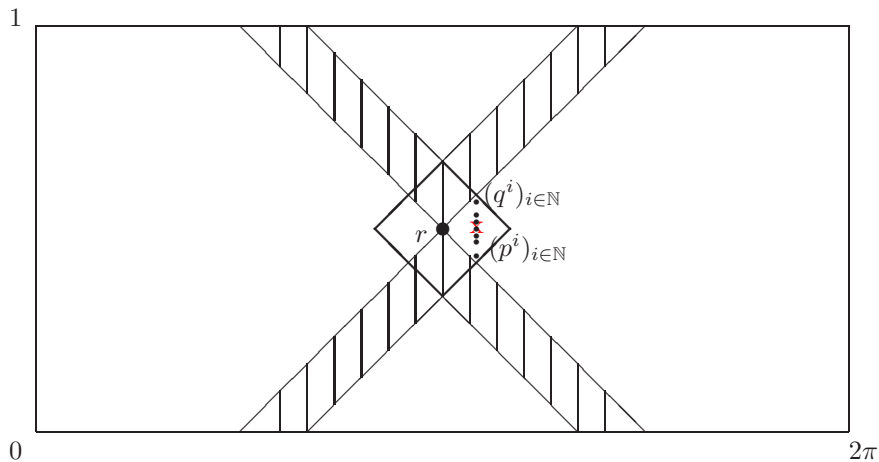
\begin{figure}[h]
\begin{center}
  \setlength{\unitlength}{1.8cm}
\begin{picture}(8,5)

\put(1,1){\line(1,0){6}}
\put(1,1){\line(0,1){3}}
\put(1,4){\line(1,0){6}}
\put(7,1){\line(0,1){3}}
\put(0.8,0.8){$0$}
\put(7,0.8){$2\pi$}
\put(0.8,4){$1$}

\put(2.5,1){\line(1,1){3}}
\put(2.5,4){\line(1,-1){3}}
\put(4,2.5){\circle*{0.1}}
\put(3.8,2.4){$r$}
\put(3,1){\line(1,1){1}}
\put(5,1){\line(-1,1){1}}
\put(3,4){\line(1,-1){1}}
\put(5,4){\line(-1,-1){1}}

\multiput(0,0)(0.2,0.2){6}{\put(3,1){\line(0,1){0.5}}}
\put(2.8,1){\line(0,1){0.3}}
\multiput(0,0)(-0.2,0.2){5}{\put(5,1){\line(0,1){0.5}}}
\put(5.2,1){\line(0,1){0.3}}
\multiput(0,0)(0.2,-0.2){6}{\put(3,4){\line(0,-1){0.5}}}
\put(2.8,4){\line(0,-1){0.3}}
\multiput(0,0)(-0.2,-0.2){5}{\put(5,4){\line(0,-1){0.5}}}
\put(5.2,4){\line(0,-1){0.3}}

\thicklines
\put(4,3){\line(1,-1){0.5}}
\put(4,3){\line(-1,-1){0.5}}
\put(4,2){\line(1,1){0.5}}
\put(4,2){\line(-1,1){0.5}}

\put(4.2,2.47){\Red{x}}
\put(4.3,2.7){$(q^i)_{i\in\mathbb{N}}$}
\put(4.25,2.7){\circle*{0.04}}
\put(4.25,2.6){\circle*{0.04}}
\put(4.25,2.55){\circle*{0.04}}
\put(4.25,2.5){\circle*{0.04}}
\put(4.34,2.3){$(p^i)_{i\in\mathbb{N}}$}
\put(4.25,2.3){\circle*{0.04}}
\put(4.25,2.4){\circle*{0.04}}
\put(4.25,2.45){\circle*{0.04}}

\end{picture}
\caption{Equivalent sequences converging to $r$ are not $T_{0}$ equivalent to $r$.}
\label{fig5}
\end{center}
\end{figure}

Consider the sequence $((0,1+ \pi + \frac{1}{n+1}) )_{ n \in \mathbb{N}}$. Surprisingly, \emph{every} point in the degenerate strip is a limit point of this sequence in the Alexandrov topology!  However, only the tip is the \emph{unique}, strong limit point.  As proven in \cite{Noldus} Theorem 8, $d$ is continuous in the \emph{strong}, but not in the Alexandrov topology.  This tells us that the strong topology is \emph{more suitable} since one would at least like $d$ to be continuous on ``the'' limit space.  Another example of the rather pathological behaviour of the Alexandrov topology is the following.  Removing the tip from the previous limit space, one obtains a timelike complete, non strongly compact (but compact in the Alexandrov topology) limit space with respect to mappings $\psi_{\delta}^{\epsilon}$ and $\zeta_{\epsilon}^{\delta}$ defined as follows: $\zeta_{\epsilon}^{\delta}$ is simply the identity,  $\psi_{\delta}^{\epsilon}$ however is defined by making a cut at $t = 1 + \pi$; $\psi_{\delta}^{\epsilon} (x,t) = (x,t) $ for $t \in \left[ 0 , 1 \right] \cup \left( 1 + \pi , 8 \right]$ and $( x , 1 + (1 - \epsilon)(t - 1) )$ for $t \in \left[ 1 , 1 + \pi \right]$.  Notice also that in the strong topology, the (first) limit space is compact and Hausdorff.     $\square$  \\* 
\\*
These last examples made clear that the strong metric gives rise to a suitable topology on $\overline{T_{0}\mathcal{S}}$.  Moreover, as proven in \cite{Noldus}, Theorem 8, the strong topology coincides with the manifold topology on a compact globally hyperbolic interpolating spacetime\footnote{As a matter of fact, for a distinguishing spacetime (with or without boundary) with finite timelike diameter, the strong topology is finer as the manifold topology \cite{Meyer} \cite{Kriele}.}.
The figures in the next example are illustrations of the remark following the definition of the timelike continuum. \\*
\\*
\textbf{Example 4} \\*
Fig.\ \ref{fig4} shows future timelike $(p^{i})_{i \in \mathbb{N}}$ and past timelike Cauchy sequences $(q^{i})_{i \in \mathbb{N}}$ which converge to $r$ in the Alexandrov topology, but $r$, $(p^{i})_{i \in \mathbb{N}}$ and $(q^{i})_{i \in \mathbb{N}}$ are not $T_{0}$ equivalent. 
Fig.\ \ref{fig5} shows that two equivalent sequences $(p^{i})_{i \in \mathbb{N}} \sim (q^{i})_{i \in \mathbb{N}}$ which converge to $r$ are not necessarily $T_{0}$ equivalent with $r$.
\\*
The ``universe'' in these pictures is $(S^{1} \times \left[0 ,1\right],d)$, where $d$ is the limit distance defined by a sequence $d_{\epsilon}$.  $d_{\epsilon}$ is constructed from $g_{\epsilon} = \Omega_{\epsilon}^{2}(t) ( -dt^{2} + d\theta^{2})$ where the smooth conformal factor $\Omega_{\epsilon}$ goes, proportionally to $\epsilon$, to zero on the shaded area $\mathcal{D}$ and to $1$ elsewhere.  Hence, $d$ is degenerate on $\mathcal{D}$, but the timelike relations between points $p \in \mathcal{D}$ and $q \in S^{1} \times \left[0 ,1\right] \setminus \mathcal{D}$ are the ones induced by $ds^{2} = -dt^{2} + d\theta^{2}$.
$\square$ 
\\*
\\*
We shall be mainly interested in the strong topology, but first I finish with stating a few properties of $d$ in the Alexandrov topology.  One can show that $d$ is continuous on the timelike continuum of $\overline{T_{0}\mathcal{S}}$ and that the Alexandrov topology has the $T_{2}$ property on $\mathcal{TCON}$, a proof can be found in \cite{Noldus1}.  Also, one can prove that $(\overline{T_{0}\mathcal{S}},d)$ is a limit space of the sequence $(\mathcal{M}_{i},g_{i})$, see \cite{Noldus1}. \\* \\*
Let us summarize our preliminary results: examples $2$ and $3$ show that we have to allow degenerate metrics and that, moreover, the Alexandrov topology has bad separation properties on the ``degenerate area''.  The afore mentioned results show that the candidate limit space has the required behaviour on the timelike continuum.  However, by a judicious choice of mappings $\psi$ and $\zeta$, one can give examples where $\overline{T_{0}\mathcal{S}}$ is \emph{not} compact in the Alexandrov topology while $\overline{T_{0}\mathcal{S}}$ is compact in the Alexandrov topology for another set mappings, see \cite{Noldus1}!   All this shows, in my opinion, that the Alexandrov topology is not sufficient and I shall concentrate on the strong topology from now on. \\* \\*
The beautiful thing about the strong topology is that it is a metric topology, and the immediate natural question which emerges is if modified Gromov-Hausdorff convergence of the sequence $(\mathcal{M}_{i},g_{i})$ forces Gromov-Hausdorff convergence of the compact metric spaces $(\mathcal{M}_{i}, D_{\mathcal{M}_{i}})$.  Recall that the Gromov-Hausdorff distance between metric spaces $(X,d_{X})$ and $(Y,d_{Y})$ is defined as $$d_{GH} ( (X,d_{X}),(Y,d_{Y})) = \inf \{ d_{H} (X,Y) | \textrm{all admissible metrics } d \textrm{ on } X \cup Y \}$$   
where a metric $d$ on the disjoint union $X \cup Y$ is \emph{admissible} iff the restriction of $d$ to $X$ and $Y$ equal $d_{X}$ and $d_{Y}$ respectively.  By $d_{H}$, I denote the Hausdorff distance associated to $d$.  
\begin{theo}
Let $(\mathcal{M},g)$ and $(\mathcal{N},h)$ be $(\epsilon , \delta)$ Lorentzian Gromov-Hausdorff close, then $d_{GH}((\mathcal{M},D_{\mathcal{M}}),(\mathcal{N}, D_{\mathcal{N}})) \leq \epsilon + \frac{3 \delta}{2}$.
\end{theo}
\textsl{Proof}:  \\*
Let $\psi : \mathcal{M} \rightarrow \mathcal{N}$ and $\zeta : \mathcal{N} \rightarrow \mathcal{M}$ be mappings which make $(\mathcal{M},g)$ and $(\mathcal{N},h)$, $(\epsilon , \delta)$ close.  Then, using that $D_{\mathcal{M}} (\zeta \circ \psi (p) , p) , D_{\mathcal{N}} (\psi \circ \zeta (q) ,q ) < \delta$, it is not difficult to derive that 
\begin{equation} \label{imp1} \left| D_{\mathcal{N}} ( \psi (p) , \psi (q)) - D_{\mathcal{M}} (p,q ) \right| < 2( \epsilon + \delta) \quad \forall p,q \in \mathcal{M} \end{equation} 
and 
\begin{equation} \label{imp2} \left| D_{\mathcal{M}} (\zeta (p) , \zeta(q) ) - D_{\mathcal{N}} (p,q) \right| < 2( \epsilon +\delta) \quad \forall p,q \in \mathcal{N}. \end{equation}
I define an admissible metric $D$ on $\mathcal{M} \cup \mathcal{N}$ by declaring that  
\begin{eqnarray*} D(p,q)& = & \min_{r \in \mathcal{M}, s \in \mathcal{N}} \frac{1}{2} \left( D_{\mathcal{M}} (p,r) + D_{\mathcal{N}} ( \psi (r), q) + D_{\mathcal{N}} (q , s ) + D_{\mathcal{M}} (\zeta (s) ,p) \right) + \\ & & (\epsilon + \delta) \end{eqnarray*}
for all $p \in \mathcal{M}$ and $q \in \mathcal{N}$, and symmetrically.  It is necessary to check that $D$ satisfies the triangle inequality.  Let $p_{1}, p_{2} \in \mathcal{M}$ and $q \in \mathcal{N}$, then
\begin{eqnarray*}
D(p_{1} , p_{2}) & \leq & \frac{1}{2} ( D_{\mathcal{M}} ( p_{1} , r_{1}) + D_{\mathcal{M}}(r_{1} , r_{2}) + D_{\mathcal{M}} (r_{2} , p_{2}) + D_{\mathcal{M}} ( p_{1} , \zeta (s_{1}) ) + \\ & & D_{\mathcal{M}} (\zeta (s_{1}) , \zeta (s_{2}) ) + D_{\mathcal{M}} ( \zeta (s_{2}) , p_{2} ) ) \\
& \leq & \frac{1}{2} ( D_{\mathcal{M}} ( p_{1} , r_{1}) + D_{\mathcal{N}} ( \psi (r_{1}) , \psi (r_{2}) ) + D_{\mathcal{M}} (r_{2} , p_{2}) + 2(\epsilon + \delta) + \\ & & D_{\mathcal{N}} (s_{1} , s_{2}) + D_{\mathcal{M}} ( \zeta (s_{2}) , p_{2} ) ) + D_{\mathcal{M}} ( p_{1} , \zeta (s_{1}) ) \\
& \leq & \frac{1}{2} ( D_{\mathcal{M}} ( p_{1} , r_{1}) + D_{\mathcal{N}} ( \psi (r_{1}), q) + D_{\mathcal{N}} (s_{1} , q) + D_{\mathcal{M}}(\zeta (s_{1}) , p_{1} )) + \\ & &  \frac{1}{2} (D_{\mathcal{M}}( p_{2} , r_{2} ) + D_{\mathcal{N}} ( \psi (r_{2}) , q ) + D_{\mathcal{N}} (q , s_{2} ) + D_{\mathcal{M}} ( \zeta (s_{2}) , p_{2}) ) + \\ & & 2(\epsilon + \delta)   
\end{eqnarray*} 
for all $r_{1},r_{2} \in \mathcal{M}$ and $s_{1} ,s_{2} \in \mathcal{N}$.  Hence
$$D(p_{1} , p_{2} ) \leq D(p_{1} ,q) + D(q , p_{2})$$
The other triangle inequalities are proven similary.  Obviously, $$D(p ,\psi(p)), D(q, \zeta (q)) \leq \epsilon + \frac{3 \delta}{2},$$ which proves the claim.   $\square$ 
\\*
\\*
Theorem $2$ reveals that any compact limit space (in the strong topology), $(\mathcal{M}^{\textrm{str}},d)$, of the modified Gromov-Hausdorff sequence $(\mathcal{M}_{i} ,g_{i})_{i \in \mathbb{N}}$ must be isometric, w.r.t. $D_{\mathcal{M}^{\textrm{str}}}$, to the limit space of the Gromov-Hausdorff sequence $(\mathcal{M}_{i} , D_{\mathcal{M}_{i}})_{i \in \mathbb{N}}$ due to the well known result of Gromov \cite{Gromov},\cite{Petersen}.  One could now proceed as before and define $\mathcal{M}^{\textrm{str}}$ by a completion procedure from the $T_{0}$ quotient of $\mathcal{S}$.  On the other hand, we can, inspired by the previous theorem, construct in a direct way a compact limit space by using the classical Gromov construction.  The reader will easily see that the $T_{0}$ quotient of $\mathcal{S}$ is dense in $\mathcal{M}^{\textrm{str}}$ in the strong topology defined by $D_{\mathcal{M}^{\textrm{str}}}$.  
\begin{theo}
The Gromov-Hausdorff limit space of the sequence $(\mathcal{M}_{i},D_{\mathcal{M}_{i}})_{i \in \mathbb{N}}$  with a suitably defined Lorentzian distance $d$, is a limit space of the sequence $(\mathcal{M}_{i} ,g_{i})_{i \in \mathbb{N}}$.
\end{theo}   
\textsl{Proof}:  \\*
Let $\psi_{i+1}^{i} : \mathcal{M}_{i} \rightarrow \mathcal{M}_{i+1}$, $\zeta^{i+1}_{i} : \mathcal{M}_{i+1} \rightarrow \mathcal{M}_{i}$ be as before and denote by $D_{i,i+1}$ the admissible metric on $\mathcal{M}_{i} \cup \mathcal{M}_{i+1}$ constructed from $\psi_{i+1}^{i}, \zeta^{i+1}_{i}$ and $D_{\mathcal{M}_{i}}, D_{\mathcal{M}_{i+1}}$ as in the proof of theorem $2$.  \\* Then, $D_{i,i+1} (p_{i} , \psi^{i}_{i+1} (p_{i})), D_{i,i+1} ( p_{i+1} , \zeta^{i+1}_{i} ( p_{i+1} )) \leq \frac{5}{2^{i+1}}$.  The following inequality is crucial:
\begin{eqnarray*}    
\left| d_{g_{i+1}} (p_{i+1} ,q_{i+1}) - d_{g_{i}}(p_{i} ,q_{i}) \right| & \leq & \frac{1}{2^{i}} + \left| d_{g_{i+1}} ( p_{i+1} , q_{i+1}) - d_{g_{i+1}} ( \psi^{i}_{i+1} ( p_{i}) , \psi^{i}_{i+1} ( q_{i})) \right| \\ & \leq &
\frac{1}{2^{i}} + D_{\mathcal{M}_{i+1}} ( p_{i+1} , \psi^{i}_{i+1} (p_{i})) + D_{\mathcal{M}_{i+1}} ( q_{i+1}, \psi^{i}_{i+1} ( q_{i})) \\ & \leq & \frac{3}{2^{i-1}} + D_{i,i+1} ( p_{i} , p_{i+1}) + D_{i ,i+1} (q_{i},q_{i+1}) 
\end{eqnarray*}
Let $\bigcup_{i \in \mathbb{N}} \mathcal{M}_{i}$ be the disjoint union of the $\mathcal{M}_{i}$ and define a metric $D$ on it by declaring that for all $i,k >0$
$$ D(p_{i} , p_{i+k}) = \min_{\left\{p_{i+j} \in \mathcal{M}_{i+j}, j=1 \ldots k-1 \right\} } \left\{ \sum_{j=0}^{k-1} D_{i+j, i+j+1}(p_{i+j} , p_{i+j+1} ) \right\} $$
Obviously, $D(p_{i} , \psi^{i}_{i+k} (p_{i})),D(p_{i+k} , \zeta^{i+k}_{i} (p_{i+k})) \leq \frac{5}{2^{i}}$ and   
$$ \left| d_{g_{i+k}} (p_{i+k} ,q_{i+k}) - d_{g_{i}}(p_{i} ,q_{i}) \right|  \leq \frac{3}{2^{i-2}} + D(p_{i},p_{i+k}) + D(q_{i},q_{i+k}) $$
I construct now the limit space as the ``boundary'' of the completion of $\left( \bigcup_{i \in \mathbb{N}} \mathcal{M}_{i} , D \right)$.  
Define 
$$\widehat{\mathcal{M}} = \left\{ (p_{i})_{i \in \mathbb{N}} | p_{i} \in \mathcal{M}_{i} \textrm{ and } D(p_{i} ,p_{j}) \rightarrow 0 \textrm{ for } i,j \rightarrow \infty \right\} $$
$\widehat{\mathcal{M}}$ has a pseudometric defined on it 
$$ D((p_{i})_{i \in \mathbb{N}} , (q_{i})_{i \in \mathbb{N}}) = \lim_{i \rightarrow \infty} D(p_{i}, q_{i}) $$
and considering the above estimates, the following Lorentz metric
$$ d((p_{i})_{i \in \mathbb{N}} , (q_{i})_{i \in \mathbb{N}}) = \lim_{i \rightarrow \infty} d_{g_{i}}(p_{i} ,q_{i})$$  
is also well defined.  I show that $D_{\widehat{\mathcal{M}}}$, defined by $d$ as $$D_{\widehat{\mathcal{M}}} ( (p_{i})_{i \in \mathbb{N}} , (q_{i})_{i \in \mathbb{N}}) = \sup_{r \in \widehat{\mathcal{M}}} \left| d((p_{i})_{i \in \mathbb{N}} , r) + d(r , (p_{i})_{i \in \mathbb{N}}) - d((q_{i})_{i \in \mathbb{N}} , r) - d(r , (q_{i})_{i \in \mathbb{N}}) \right| $$ equals $D$ on $\widehat{\mathcal{M}}$.  Suppose there exist sequences $(p_{i})_{i \in \mathbb{N}}$, $(q_{i})_{i \in \mathbb{N}}$ and $\delta >0$ such that $$D_{\widehat{\mathcal{M}}} ((p_{i})_{i \in \mathbb{N}}, (q_{i})_{i \in \mathbb{N}}) > D((p_{i})_{i \in \mathbb{N}} ,(q_{i})_{i \in \mathbb{N}}) + \delta ;$$
then there exists a sequence $(r_{i})_{i \in \mathbb{N}}$, such that for $k$ big enough:
$$ \left| d_{g_{k}} (p_{k} ,r_{k} ) + d_{g_{k}} (r_{k} , p_{k}) - d_{g_{k}} (q_{k} ,r_{k}) - d_{g_{k}}(r_{k} ,q_{k}) \right| > D_{\mathcal{M}_{k}} ( p_{k} ,q_{k}) + \frac{\delta}{2},$$ which is impossible by definition of $D_{\mathcal{M}_{k}}$.  Hence, suppose that there exist sequences $(p_{i})_{i \in \mathbb{N}}$, $(q_{i})_{i \in \mathbb{N}}$, $\delta >0$ such that $$D_{\widehat{\mathcal{M}}} ((p_{i})_{i \in \mathbb{N}}, (q_{i})_{i \in \mathbb{N}}) + \delta < D((p_{i})_{i \in \mathbb{N}} ,(q_{i})_{i \in \mathbb{N}})$$
Choose $k > \frac{ ln(\frac{176}{\delta})}{ln(2)}$ big enough such that 
$$ \left| D((p_{i})_{i \in \mathbb{N}} ,(q_{i})_{i \in \mathbb{N}}) - D_{\mathcal{M}_{k}}(p_{k} ,q_{k}) \right| < \frac{\delta}{4}  $$
and $D(p_{k}, (p_{i})_{i \in \mathbb{N}}), D(q_{k} , (q_{i})_{i \in \mathbb{N}}) < \frac{5}{2^{k}}$; then it is not difficult to see that the hypothesis implies that
$$ \left| d_{g_{k}} ( p_{k} , r_{k}) + d_{g_{k}} (r_{k} ,p_{k}) - d_{g_{k}} (q_{k} ,r_{k}) - d_{g_{k}}(r_{k} ,q_{k})\right| + \frac{\delta}{2} < D_{\mathcal{M}_{k}} (p_{k} ,q_{k}) $$ for all $r_{k} \in \mathcal{M}_{k}$, which is impossible and therefore $D_{\widehat{\mathcal{M}}} = D$.    \\*
Hence, $(\widehat{\mathcal{M}},d)$ is a compact limit space in the strong topology since $\widehat{\mathcal{M}}$ is compact with respect to $D$ (it is a good exercise for the reader to check this).  I claim now that the $T_{0}$ quotient of $(\widehat{\mathcal{M}},d)$ is the desired limit space $(\mathcal{M}^{\textrm{str}} ,d)$.  This is an immediate consequence of the fact that the Gromov-Hausdorff distance between $(\mathcal{M}^{\textrm{str}} ,D)$ and $(\mathcal{M}_{k} , D_{\mathcal{M}_{k}})$ is less than $\frac{5}{2^{k+1}}$ and the inequality
\begin{equation} \label{imp3} \left| d((p_{i})_{i \in \mathbb{N}} ,(q_{i})_{i \in \mathbb{N}}) - d_{g_{k}} (r_{k} ,s_{k}) \right| \leq \frac{3}{2^{k-2}} + D((p_{i})_{i \in \mathbb{N}} , r_{k}) + D((q_{i})_{i \in \mathbb{N}} , s_{k}) \end{equation}   $\square$ \\*
\\*  
The uniqueness of the limit space is easily proven from theorem $2$.  In fact, we have the following result.
\begin{theo}
Let $(\mathcal{M}_{1},d_{1})$, $(\mathcal{M}_{2},d_{2})$ be pairs, where $\mathcal{M}_{i}$ is a set with a Lorentz distance $d_{i}$ defined on it, such that $\mathcal{M}_{i}$ is compact in the strong metric topology defined by $d_{i}$ for $i=1,2$.  Then $(\mathcal{M}_{1},d_{1})$ and  $(\mathcal{M}_{2},d_{2})$ cannot be distinguished by the Gromov-Hausdorff uniformity iff they are isometric.
\end{theo}
\textsl{Proof}: \\*
Let $\psi_{n} : \mathcal{M}_{1} \rightarrow \mathcal{M}_{2}$, $\zeta_{n} : \mathcal{M}_{2} \rightarrow \mathcal{M}_{1}$ such that $\psi_{n}$ and $\zeta_{n}$ make $(\mathcal{M}_{1},d_{1})$ and $(\mathcal{M}_{2},d_{2})$, $(\frac{1}{n} , \frac{1}{n})$ close.  Then, the inequalities (\ref{imp1}, \ref{imp2}) reveal that 
$$ \left| D_{\mathcal{M}_{2}} ( \psi_{n} (p) , \psi_{n} (q)) - D_{\mathcal{M}_{1}} (p,q ) \right| < \frac{4}{n} \, \forall p,q \in \mathcal{M}_{1} $$
and
$$ \left| D_{\mathcal{M}_{1}} ( \zeta_{n} (p) , \zeta_{n} (q)) - D_{\mathcal{M}_{2}} (p,q ) \right| < \frac{4}{n} \, \forall p,q \in \mathcal{M}_{2}. $$
This, combined with,
$$ D_{\mathcal{M}_{1}} ( \zeta_{n} \circ \psi_{n} (p) , p), D_{\mathcal{M}_{2}}( \psi_{n} \circ \zeta_{n} (q) ,q) < \frac{1}{n} \, \forall p \in \mathcal{M}_{1}, q \in \mathcal{M}_{2} $$
implies that we can find a subsequence $(n_{k})_{k \in \mathbb{N}}$ and an isometry $\psi$ such that 
$$\psi_{n_{k}} \stackrel{k \rightarrow \infty}{\rightarrow} \psi$$ and $$\zeta_{n_{k}} \stackrel{k \rightarrow \infty}{\rightarrow} \psi^{-1}$$ pointwise (see Appendix C).  Clearly, $\psi$ must preserve the Lorentz distance.  $\square$
  
\section{Some first properties of the limit space}
Now, I shall study some first properties of the chronological relation and start working towards a good definition of the causal relation $\leq$.  I will end this section by giving some conditions on the spacetimes $(\mathcal{M}_{i} ,g_{i})$ which imply that the timelike continuum of the limit space $(\mathcal{M}^{str},d)$ is ``as large as possible''.   To start with, I give an example as a warm-up for the phenomena we need to consider. \\*
\\*
\textbf{Example 5}
Choose $T, \epsilon >0$ and consider the cylinder $S^{1} \times \left[-T,T \right]$ with the usual coordinates $(\theta ,t)$.  Let $f$ be the function constructed in example $2$ and redefine $\beta_{\epsilon}$ as $\beta_{\epsilon} (t) = \epsilon + ( 1 - \epsilon) \chi( t - \frac{T}{6} )$.  Define $\rho_{\epsilon}$ as 
$$ \rho_{\epsilon}(t,\theta) = \frac{\beta_{\epsilon}( \cdot +  \frac{T}{3})*f( \cdot + \frac{T}{6})f(- \cdot + \frac{T}{6})}{\int_{-\infty}^{+\infty}f(x+ \frac{T}{6})f(-x+\frac{T}{6})dx} (t)$$
$\rho_{\epsilon}$ is a function which equals $\epsilon$ for $t \leq - \frac{T}{3}$, $1$ for $t \geq 0$ and undergoes a smooth, injective transition in the interval $\left[ - \frac{T}{3} , 0 \right]$.  Define metric tensors
$$ g_{\epsilon}(t, \theta) = -dt^{2} + \rho_{\epsilon}^{2}(t)d\theta^{2} $$  
Denote by $d_{\epsilon}$ the associated Lorentz distances, set $\psi_{\delta}^{\epsilon}, \zeta_{\epsilon}^{\delta}$ equal to the identity on $S^{1} \times \left[ -T, T \right]$, and remember that $d$ is the ``limit distance'', $d = \lim_{\epsilon \rightarrow 0} d_{\epsilon}$.  What does the ($T_{0}$ quotient of) the limit space looks like?   Let me give a dynamical picture of what happens: for any $\epsilon > 0$, the spacetime at hand is a tube of radius $\epsilon$ for $t \leq - \frac{T}{3}$ and $1$ for $t \geq 0$.  In the limit $\epsilon \rightarrow 0$, we are left with a one dimensional timelike line $\left[ -T ,-\frac{T}{3} \right]$ and a cylinder $S^{1} \times \left[0 , T \right]$ of radius one which are connected between $-\frac{T}{3}$ and $0$ by a tube. In this example, we close $\ll$ to $\leq$ by defining $J^{\pm}(p)$ as $J^{\pm}(p) = \overline{I^{\pm} (p)}$.  Obviously $J^{+}$ is the dual of $J^{-}$, and $\leq$ is a reflexive, antisymmetric and transitive relation.  Moreover, for $r \in \left(-T, -\frac{T}{3} \right)$, one has that $I^{+}(r) \neq \bigcap_{q \leq r, q \neq r} I^{+} (q)$ since $J^{\pm}(r) = I^{\pm}(r) \cup \left\{ r \right\}$.  $\square$               
\\*
\\*
The reader should try to make the following exercises, before watching examples $6$ and $7$, in order to understand what does \emph{not} work. \\*
\\*
\textbf{Exercises}  \\*
Define the timelike continuum, $\mathcal{TCON}$ of $\mathcal{M}^{\textrm{str}}$, as before.
\begin{itemize}
\item Construct a limit space with an element $r \in \mathcal{TCON}$ such that $\overline{I^{+} (r)} \neq \bigcap_{s \ll r} I^{+} (s)$, i.e., $(\mathcal{M}^{str},d)$ is not causally continuous\footnote{By this I mean that $p \rightarrow I^{\pm}(p)$ is not outer continuous in the usual $C^{0}$ topology.}.  Hence, define the causal future $J^{+}(r)$ of $r \in \mathcal{TCON}$ as $J^{+}(r) = \bigcap_{s \ll r} I^{+} (s)$; the  causal past is defined dually.  
\item Construct a limit space with two points $p,q$ such that $I^{-}(p) = I^{-}(q)$, $I^{+}(p) \subsetneq I^{+}(q)$, but $q$ cannot be in the causal past of $p$ without breaking inner continuity of $r \rightarrow J^{-}(r)$ in the usual $C^{0}$ topology.
\item Is the timelike continuum open?
\item Prove that on $\mathcal{TCON}$ the Alexandrov and strong topology coincide. 
\end{itemize}  $\square$
\\*
\\*
\textbf{Example 6} \\*
In this example, I construct a solution to the first exercise.  The second one is considerably easier and is not treated.  Take the ``cylinder universe'' $\mathcal{CYL} = (S^{1} \times \left[0,1 \right], -dt^{2} + d\theta^{2})$ and define $(\mathcal{M}_{\epsilon}, -dt^{2} + d\theta^{2})$ by removing \\* $D^{+}(K^{+}((\pi , \frac{3}{4}))) \setminus K^{+}((\pi , \frac{3}{4}))$ from $\mathcal{CYL}$.  Recall that $K^{+}((\pi , \frac{3}{4}))$ is the future sphere of radius $\epsilon$ around $(\pi , \frac{3}{4})$ and $D^{+}(A)$ is the domain of dependence of a partial Cauchy surface $A$.  It is easy to see that $(\left(S^{1} \times \left[0,1\right]\right) \setminus I^{+}((\pi , \frac{3}{4})) , d)$ is a strong limit space of the sequence $(\mathcal{M}_{\epsilon}, -dt^{2} + d\theta^{2})_{\epsilon}$ where $d$ is the usual Lorentz distance associated to the metric tensor $-dt^{2} + d\theta^{2}$ on $\mathcal{M}^{str}$.  One can see that $(\pi - \frac{1}{4} , 1) \notin \overline{I^{+}((\pi + \frac{1}{4}, \frac{1}{2}))}$ but $(\pi - \frac{1}{4} , 1) \in \bigcap_{q \ll (\pi + \frac{1}{4}, \frac{1}{2})} I^{+}(q)$. $\square$
\\*
\\*
\textbf{Example 7} \\*
I give an example in which the timelike continuum is not open.  As in the previous example, I consider the cylinder universe.  For any $n \geq 6$, denote by $m(n)$ the largest even  integer (just for convenience) such that $\frac{1}{m(n)(m(n) + 1)} \geq \frac{3}{n}$.  Define a conformal factor $\Omega_{n}$ as follows:
\begin{itemize}
\item $\Omega_{n}(t) = 1$ for $t \geq \frac{1}{2} + \frac{1}{n}$
\item for all $0 < k < m(n)$, $k$ even, $\Omega_{n}(t) = \frac{1}{n}$ for $t \in \left[\frac{1}{k+1} + \frac{1}{n} ,  \frac{1}{k} - \frac{1}{n} \right]$ and $\Omega_{n}$ smoothly increases from $\frac{1}{n}$ to $1$ on the interval $\left[ \frac{1}{k} - \frac{1}{n} , \frac{1}{k} + \frac{1}{n} \right]$.  
\item for all $1 < k < m(n)$, $k$ odd, $\Omega_{n}(t) = 1$ for $t \in \left[\frac{1}{k+1} + \frac{1}{n} ,  \frac{1}{k} - \frac{1}{n} \right]$ and $\Omega_{n}$ smoothly decreases from $1$ to $\frac{1}{n}$ on the interval $\left[ \frac{1}{k} -\frac{1}{n} , \frac{1}{k} + \frac{1}{n} \right]$.
\item for $t \leq \frac{1}{m(n)} + \frac{1}{n}$, $\Omega_{n} (t) = 1$
\end{itemize}       
Consider the sequence of spacetimes $(S^{1} \times \left[-1,1\right], \Omega_{n}^{2}(t) ( -dt^{2} + d\theta^{2}))_{n \geq 6 }$, and the associated strong limit space $(\mathcal{M}^{str} ,d)$ (w.r.t. mappings which equal the identity) then any point $p$ with time coordinate $0$ belongs to the timelike continuum.  However any neighborhood of such $p$ contains points which do not belong to the timelike continuum. $\square$ 
\\*
\\*
I show that $J^{+}(r)$ is closed for $r \in \mathcal{TCON}$ (the dual statement follows identically).  Remark first that for $q \ll r$, $\overline{I^{+}(r)} \subset I^{+}(q)$ which follows immediately from the continuity of $d$ and the reversed triangle inequality.  Hence, $\overline{I^{+}(r)} \subset J^{+}(r)$.  Let $(p^{i})_{i \in \mathbb{N}}$ be a sequence in $J^{+}(r)$ converging to an element $p$ and choose $q \ll r$.  Pick $s$ such that $q \ll s \ll r$, then $p \in \overline{I^{+}(s)}$ and hence, $p \in I^{+}(q)$, which finishes the proof. \\*
\\*
One might wonder if two chronologically related points can be joined by a timelike geodesic.  In the next example, I show that one can at most expect such geodesic to be causal.
\\*
\\*
\textbf{Example 8} \\*
I construct a limit space $\mathcal{M}^{\textrm{str}}$ in which the unique geodesic (causal curve with longest length) between two points $p \ll q$ is broken into a timelike and null part.  The idea is to construct conformal factors $\Omega_{\epsilon}$ which equal $100$ in a specific area $A$ containing $q$, but not $p$, and $1$ outside $A$ apart from a small tube within a distance $\epsilon$ (w.r.t. the obvious ``Riemannian'' metric) from $A$ such that the geodesics from $p$ to $q$ bend towards the lightcone of $p$ as $\epsilon$ goes to zero since they favour travelling inside $A$ as much as possible.  We start from the cylinder universe $\mathcal{CYL}$ and let $p = (\pi ,\frac{1}{4})$, $q = (\pi , \frac{3}{4})$.  The past sphere of radius $\frac{1}{4}$ around $q$ intersects the null geodesic $t + \theta = \frac{1}{4} + \pi$ trough $p$ in the point $r = ( \pi - \frac{3}{16} ,  \frac{7}{16})$.  Consider the null ray $t - \theta = \frac{5}{8} - \pi$ trough $r$, then let $B \subset \left\{(\theta,t) | t-\theta \geq \frac{5}{8} - \pi \right\}$ be a closed rectangle containing $A(p,q) \cap \left\{(\theta,t) | t-\theta \geq \frac{5}{8} - \pi \right\}$ in its interior such that the distance of $B$ to the boundaries of $\mathcal{CYL}$, with respect to the usual euclidian metric $\tilde{D}$ defined by $dt^{2} + d\theta^{2}$, is greater than zero.  Define $A$ by suitably rounding off the corners of $B$ outside $A(p,q)$.  We define now the outward $\epsilon$-tube $A^{\epsilon}$ of $A$  as the set of all $x \notin A$ which are a distance less or equal than $\epsilon$ apart from $A$, i.e.,
$$ A^{\epsilon} = \left\{ x \notin A | \exists a \in A : \tilde{D}(x,a) \leq \epsilon \right\} .$$
It is well known that for $\epsilon$ sufficiently small $\partial A^{\epsilon} \setminus \partial A$ is smooth if $\partial A$ is.  For $\epsilon$ sufficiently small, let $\Omega_{\epsilon}$ be a function which equals $100$ on $A$, $1$ in the complement of $A^{\epsilon} \cup A$ and undergoes a transition in $A^{\epsilon}$ which is only dependent on, and decreasing in, the radial outward coordinate.  As usual $g_{\epsilon} = \Omega^{2}_{\epsilon}(t,x) ( -dt^{2} + d\theta^{2} )$, $\psi^{\epsilon}_{\delta}$ and $\zeta^{\delta}_{\epsilon}$ equal the identity, and $d = \lim_{\epsilon \rightarrow 0} d_{\epsilon}$.  Define $\gamma_{1} : \left[0,1\right] \rightarrow \mathcal{M}^{\textrm{str}}$ as $\gamma_{1}(t) = (\pi - \frac{3 t}{16} , \frac{4 + 3t}{16})$. Define $\gamma_{2} : \left[ 0 , 1 \right] \rightarrow \mathcal{M}^{str}$ as the line running from $(\pi - \frac{3}{16} , \frac{7}{16})$ to $(\pi , \frac{3}{4})$ with uniform speed.  I claim that $\gamma = \gamma_{2} \circ \gamma_{1}$ is (upon reparametrization) the unique distance maximizing causal curve from $p$ to $q$ with length equal to $d(p,q)$.  With $\tilde{t} = t - \frac{1}{4}$, $\tilde{\theta} = \theta - \pi$ we have that the square of the length of any continuous causal curve which is linear from $p$ to a point $s \in \left\{(\theta,t) | t-\theta = \frac{5}{8} - \pi \right\}$ and from $s$ to $q$, equals
$$ 100^{2}\left[ \left(\tilde{t} - \frac{1}{2}\right)^{2} - \left(\tilde{t} - \frac{3}{8}\right)^{2}\right] + \left[ \tilde{t}^{2} - (\tilde{t} - \frac{3}{8})^{2} \right] = \frac{3 - 100^{2}}{4}\tilde{t} + \frac{7 \cdot 100^{2} - 9}{64}$$
for $\frac{3}{16} \leq \tilde{t} \leq \frac{7}{16}$, which proves the first part of the claim.  It is up to the reader to prove that the maximal length, $d(p,q)$, equals $25$.  $\square$                \\*
\\*    
Intuitively, one would say in the last example that $ \mathcal{M}^{\textrm{str}} \setminus \mathcal{TCON} = \partial \mathcal{M}^{\textrm{str}}$.  Hence, it is time to give an intrinsic definition of the past and future boundary.
\begin{deffie}
A sequence $(p_{i})_{i \in \mathbb{N}} \in \mathcal{M}^{\textrm{str}}$ represents a point of the \emph{past} boundary, $\partial_{\textrm{P}} \mathcal{M}^{\textrm{str}}$, iff $\forall \epsilon > 0$, $\exists i_{0} \in \mathbb{N}:$  $\forall i \geq i_{0}$ 
$$ p_{i} \in \left( \partial_{\textrm{P}} \mathcal{M}_{i} \right)^{\epsilon} $$
where, now, $$A_{i}^{\epsilon} = \left\{ q_{i} \in \mathcal{M}_{i} | \exists a_{i} \in A_{i} : D_{\mathcal{M}_{i}} ( a_{i} , q_{i}) < \epsilon\right\} $$
for all $A_{i} \subset \mathcal{M}_{i}$.  The future boundary $\partial_{\textrm{F}} \mathcal{M}^{\textrm{str}}$ is defined dually.  $\square$ 
\end{deffie}
We have the following theorem.
\begin{theo}
The past and future boundaries of $\mathcal{M}^{\textrm{str}}$ are achronal sets.
\end{theo}
\textsl{Proof}: \\*
Suppose we can find $p,q \in \partial_{\textrm{P}} \mathcal{M}^{\textrm{str}}$ such that $d(p,q) > 0$.  Let $\epsilon = \frac{d(p,q)}{2}$ and choose $i$ sufficiently large such that 
\begin{itemize}
\item $p_{i},q_{i} \in \left( \partial_{\textrm{P}} \mathcal{M}_{i} \right)^{\epsilon}$
\item $\left| d_{g_{i}} (p_{i} , q_{i}) - d(p,q) \right| < \frac{\epsilon}{2}$.
\end{itemize}
Then, by definition, there exist points $r_{i},s_{i} \in \partial_{\textrm{P}} \mathcal{M}_{i}$ such that $$D_{\mathcal{M}_{i}}(r_{i},p_{i}), D_{\mathcal{M}_{i}}(s_{i} ,q_{i}) < \epsilon$$ and, moreover, $d_{g_{i}}(p_{i},q_{i}) > \frac{3 \epsilon}{2}$.  Hence, since $\partial_{\textrm{P}} \mathcal{M}_{i}$ is spacelike, we have that
$$ \max_{t_{i} \in \partial_{\textrm{P}} \mathcal{M}_{i}} d_{g_{i}}(t_{i},p_{i}), \max_{t_{i} \in \partial_{\textrm{P}} \mathcal{M}_{i}} d_{g_{i}}(t_{i},q_{i}) < \epsilon $$
But, on the other hand, for $t_{i} \in I^{-}(p_{i}) \cap \partial_{\textrm{P}} \mathcal{M}_{i}$, we have that $d_{g_{i}} (t_{i}, q_{i}) > \frac{3 \epsilon}{2}$ which gives the necessary contradiction.  The proof for the future boundary is similar. $\square$
\\*
\\*   
Using similar arguments, it is easy to see that there exists no point to the past of $\partial_{\textrm{P}} \mathcal{M}^{\textrm{str}}$, nor does there exist a point to the future of $\partial_{\textrm{F}} \mathcal{M}^{\textrm{str}}$.   \\*
\\*
As theorems $2$, $3$ and $4$ show, the strong metric is an important object.  One might wonder if the strong metric could be a path metric for some spacetime $(\mathcal{M},g)$, since such property is stable in the Gromov-Hausdorff limit.  Moreover, a path metric reveals some interesting topological properties of the underlying space.  Unfortunately, we have the following result.  
\begin{theo}
The strong metric $D_{\mathcal{M}}$ is \emph{never} a path metric for any spacetime $(\mathcal{M},g)$. 
\end{theo} 
\textsl{Proof}: \\*
Suppose such a spacetime $(\mathcal{M},g)$ exists, and choose $p \in \partial_{\textrm{P}} \mathcal{M}$, $q \in \partial_{\textrm{F}} \mathcal{M}$ such that $d_{g}(p,q) = tdiam (\mathcal{M}) = D_{\mathcal{M}}(p,q) $.  I show that there exists no point $x$ such that $D_{\mathcal{M}} (p,x) = D_{\mathcal{M}}(x,q) = \frac{D_{\mathcal{M}}(p,q)}{2}$, which is a contradiction.  Let $r$ be a point such that $d_{g}(p,r) = d_{g}(r ,q) = \frac{d_{g}(p,q)}{2}$, then $$D_{\mathcal{M}} ( p,r ), D_{\mathcal{M}}(r,q) > \frac{d_{g}(p,q)}{2}$$  The first part is easily seen by noticing that for $s \in E^{+}(r) \setminus \left\{r \right\}$ $$\left| d_{g}(p,s) - d_{g}(r,s) \right| = d_{g}(p,s) > d_{g} (p,r).$$ The second part is proven similary.  Let $x \neq r$; then we have to distinguish two cases:
\begin{itemize}
\item $\max_{t \in \partial_{P} \mathcal{M}} d_{g}(t,x) \leq \frac{d_{g}(p,q)}{2}$
\item $\max_{t \in \partial_{P} \mathcal{M}} d_{g}(t,x) > \frac{d_{g}(p,q)}{2}$
\end{itemize} 
Suppose the former is true, then there exists a point $s \in I^{-}(r)$ which is not in the causal past of $x$.\footnote{If this were not true then $r \leq x$ which would imply that $d_{g}(p,x) > \frac{d_{g}(p,q)}{2}$ which is a contradiction.}  Hence $\left| d_{g} (s,q) - d_{g}(s,x) \right| > \frac{d_{g}(p,q)}{2}$, which implies that $D_{\mathcal{M}}(x,q) > \frac{d_{g}(p,q)}{2}$.  \\*
Suppose the latter is satisfied; then $\max_{t \in \partial_{\textrm{F}} \mathcal{M}} d_{g}(x,t) < \frac{d_{g}(p,q)}{2}$ and this case is similar to the previous one. $\square$ 
\\*
\\*  
It would be interesting to give a criterion which guarantees that $\mathcal{M}^{\textrm{str}} \setminus \partial \mathcal{M}^{\textrm{str}} = \mathcal{TCON}$ and every point of the boundary is the limit point of a timelike Cauchy sequence.  Particularly from the physical point of view, it is not entirely clear what degenerate regions would mean.  Moreover, it is far from easy to define a suitable causal relation between two points belonging to them. 
\\*
\\*
\textbf{Intermezzo}: \\*
I suggest three, at first sight different, control mechanisms which prohibit the limit space from containing ``degenerate regions''.  Let $\alpha : \mathbb{R}^{+} \rightarrow  \mathbb{R}^{+}$ be a strictly increasing, continuous function such that $\alpha (x) \leq x$ for all $x \in \mathbb{R}^{+}$.  We say that $(\mathcal{M},g)$ has the $\mathcal{C}^{+}_{\alpha}, \mathcal{C}^{-}_{\alpha}$ or $\mathcal{C}_{\alpha}$ property iff for any $tdiam(\mathcal{M}) \geq \epsilon > 0$ we have that respectively:
\begin{itemize}
\item  $\alpha (\epsilon) \leq \min_{p \in \mathcal{M}^{\downarrow \epsilon}} \left[ \max_{r \in \overline{B_{D_{\mathcal{M}}} (p , \epsilon )}}  d_{g}(p,r)  \right] \leq \epsilon$
\item $\alpha (\epsilon) \leq \min_{p \in \mathcal{M}^{\uparrow \epsilon}} \left[ \max_{r \in \overline{B_{D_{\mathcal{M}}} (p , \epsilon )}} d_{g}(r,p) \right] \leq \epsilon$
\item $\alpha (\epsilon) \leq \min_{p \in \mathcal{M}} \left[ \max_{r \in \overline{B_{D_{\mathcal{M}}} (p , \epsilon )}} \left( d_{g}(r,p) + d_{g}(p,r) \right) \right] \leq \epsilon $
\end{itemize}
where $\mathcal{M}^{\downarrow \epsilon} = \left\{ p \in \mathcal{M} | p \notin \left(\partial_{\textrm{F}} \mathcal{M} \right)^{\epsilon} \right\}$ and $\mathcal{M}^{\uparrow \epsilon} = \left\{ p \in \mathcal{M} | p \notin \left(\partial_{\textrm{P}} \mathcal{M} \right)^{\epsilon} \right\}$. \\*   
Clearly, not all functions $\alpha$ are meaningful and should satisfy $\alpha(x) + \alpha(y) \leq \alpha(x+y)$ for all $0 \leq x,y \leq x + y \leq tdiam (\mathcal{M})$.  This condition follows easily from the reverse triangle inequality satisfied by $d_{g}$ and the triangle inequality satisfied by $D_{\mathcal{M}}$.  Basic functions $\alpha_{n}^{K}$, $n >1$ and $K >0$, could be constructed by declaring that
$$ \alpha_{n}^{K} (x) = K \left(\frac{x}{K}\right)^{n} \textrm{ for } x \leq K \textrm{ and } x \textrm{ otherwise.} $$  
Obviously, $\alpha_{1} \leq \alpha_{2}$, implies that $\mathcal{C}^{\pm}_{\alpha_{2}} \subseteq \mathcal{C}^{\pm}_{\alpha_{1}}$.   Logically, $\mathcal{C}^{+}_{\alpha}$ and $\mathcal{C}^{-}_{\alpha}$ imply $\mathcal{C}_{\alpha}$, but $\mathcal{C}_{\alpha}$ implies neither of them.  The above expressions tell us there is a balance between local and global causal relations, in the sense that the local relations cannot become ``arbitrarily small'' while the global relations remain almost unaltered.  This perhaps needs a bit of explanation.  As an example, consider again the cylinder universe $\mathcal{CYL}$ and let $p = (\pi, \frac{2}{3})$.  Consider the ball $B(p , \frac{1}{100})$ of radius $\frac{1}{100}$ determined by the usual (observer dependent) Riemannian metric tensor $dt^{2} + d\theta^{2}$.  Construct a conformal factor $\Omega$ such that $\Omega(\theta ,t)$ equals $10^{-2003}$ for $(\theta ,t) \in B(p, \frac{1}{101})$, undergoes a smooth transition on the shell $\overline{B(p ,\frac{1}{100})} \setminus B(p, \frac{1}{101})$, and equals $1$ everywhere else.  It is easy to see that the strong metric determined by $(S^{1} \times \left[ 0 ,1 \right] , \Omega(\theta ,t)^{2} ( -dt^{2} + d\theta^{2} ))$ remains almost unchanged while the local Lorentzian distance changes drastically!  So the idea behind the concept is, losely speaking,  to control the conformal factor with respect to some ``reference metric''.  The following result is of main interest.
\begin{theo}
For any $\alpha$ satisfying the above conditions, the $\mathcal{C}^{+}_{\alpha}, \mathcal{C}^{-}_{\alpha}$ and $\mathcal{C}_{\alpha}$ properties are stable under generalized, Gromov-Hausdorff convergence.  This means that if, say, $((\mathcal{M}_{i},g_{i}))_{i\in \mathbb{N}}$ is a generalized, Gromov-Hausdorff Cauchy sequence such that $(\mathcal{M}_{i},g_{i}) \in \mathcal{C}^{+}_{\alpha}$ for all $i \in \mathbb{N}$, then $(\mathcal{M}^{\textrm{str}}, d) \in \mathcal{C}^{+}_{\alpha}$.
\end{theo}
The proof is left as an exercise to the reader (see Appendix B).  It is not difficult to see that $\mathcal{M} \setminus \partial \mathcal{M}^{\textrm{str}} = \mathcal{TCON}$ for a (limit) spacetime satisfying the $\mathcal{C}^{+}_{\alpha}$ and $\mathcal{C}^{-}_{\alpha}$ properties.  Moreover, every point of the boundary of such space, which does not equal $\partial_{F} \mathcal{M} \cap \partial_{P} \mathcal{M}$, is equivalent to a timelike Cauchy sequence.  There are a few serious questions which can be posed with respect to the above categories of objects. \\* \\* 
\textbf{Questions} 
\begin{itemize}
\item Can you find spacetimes such that only one of the properties $\mathcal{C}^{+}_{\alpha}, \mathcal{C}^{-}_{\alpha}$ or $\mathcal{C}_{\alpha}$ is satisfied?  If not, are some of them equivalent depending on $\alpha$?     
\item Does the limit space of a $\mathcal{C}^{+}_{\alpha}$ sequence satisfy $\mathcal{M}^{\textrm{str}} \setminus \partial \mathcal{M}^{\textrm{str}} = \mathcal{TCON}$ ?   
\item Same question for the limit space of a $\mathcal{C}^{-}_{\alpha}$ or $\mathcal{C}_{\alpha}$ sequence.
\end{itemize} $\square$ 
\\*
\\*
Clearly, in order to answer these questions, we need a deeper understanding of the strong metric.  I leave this for future investigation. $\square$
\section{Conclusions}
In this paper, I have introduced a new viewpoint on the Alexandrov topology on a compact spacetime, which distinguishes itself in the limit in case the limit space contains degenerate regions.  Theorems 2, 3 and 4 show that this metric topology is intimate involved in the Lorentzian Gromov-Hausdorff convergence mechanism, which is one of the most surprising results of this paper.  Meanwhile, concepts as the timelike continuum and timelike capacity were introduced.  While the timelike capacity is stable in the limit,  the timelike continuum is not since it is a qualitative, and not quantitative notion.  Hence, a new, quantitative control device had to be invented as to guarantee that in the limit the timelike continuum would remain ``large''.  These are the $\mathcal{C}^{\pm}_{\alpha}$ properties introduced in the intermezzo concluding section 3.  \\* 
\\*
So far, the causal relationship and questions concerning timelike (causal) curves were not treated.  In section 3, many examples made clear that answers to these questions are not obvious, and the topic is mainly left for future investigation \cite{Noldus2}.  In order to give the reader an idea what lies further ahead, I now present some open questions regarding the properties of $(\mathcal{M}^{\textrm{str}},d)$, which I shall treat in the third paper. \\*
\\*
\textbf{Questions} 
\begin{itemize}
\item Is it true that two chronologically related points can be connected by a timelike curve ? 
\item Can one suitably define $\leq$ such that Lorentzian causal path connectivity is stable? 
\item Is $\mathcal{M}^{\textrm{str}}$ path connected when all $(\mathcal{M}_{i},g_{i})$ are?
\end{itemize} 
$\square$
\\*
\\*
To whet the appetite, I invite the reader to convince himself that in order to define $\leq$, it is necessary to appeal to the notion of K-causality, which was first defined by Sorkin and Woolgar \cite{Sorkin}. \\*
\\*
The notion of the strong metric gives an observer independent scale.  Using this idea, it is possible to define a sound notion of dimension for a Lorentz space, $(\mathcal{M},d)$, which is not a spacetime.  Given $(\mathcal{M},d)$ and a  spacetime $(\mathcal{N},g)$ and $\epsilon > 0$.   A mapping $f: (\mathcal{M},d) \rightarrow (\mathcal{N},g)$ is an $\epsilon$-embedding iff 
$$ \left|d_{g}(f(p),f(q)) - d(p,q) \right| < \epsilon $$
We define $\textrm{Widim}_{\epsilon} (\mathcal{M},d)$ as the minimal number $k$ such that there exists a spacetime $(\mathcal{N},g)$ of dimension $k$ and a continuous $\epsilon$-embedding $f:(\mathcal{M},d) \rightarrow (\mathcal{N},g)$.  Clearly $\textrm{Widim}_{\epsilon} (\mathcal{M},d)$ is monotonically decreasing in $\epsilon$.  Define the injectivity radius $\textrm{inj}(f)$ of $f$ as the minimal number $\delta$ such that $D_{\mathcal{M}}(p,q) > \delta$ implies that $f(p) \neq f(q)$.   Clearly, if $f$ is an $\epsilon$-embedding, $\textrm{inj}(f) \leq 2 \epsilon$.  This is a consequence of the inequality
$$ D_{\mathcal{N}}(f(p),f(q)) \geq D_{\mathcal{M}} (p,q) - 2 \epsilon $$
Let me comment on the philosophy behind this definition.  There are strong hints that spacetime is not a manifold, but a ``structure which looks like a manifold on a scale larger than the Planck length $\ell_{p}$''.  So typically $\epsilon$ will be of order of the Planck length.  The definition says that, by eventually identifying points which are at most $2 \ell_{p}$ apart, and by perturbing the metric relations by a quantity of value smaller than the Planck length, the Lorentz space at hand is embeddable in a smooth spacetime.     
\section{Appendix A}
I prove here that the relation introduced on page $7$ is truly an equivalence relation $\sim$.  Suppose $(p^{i})_{i \in \mathbb{N}}$, $(q^{i})_{i \in \mathbb{N}}$ and $(r^{i})_{i \in \mathbb{N}}$ are future timelike Cauchy sequences such that $(p^{i})_{i \in \mathbb{N}} \sim (q^{i})_{i \in \mathbb{N}}$ and $(q^{i})_{i \in \mathbb{N}} \sim (r^{i})_{i \in \mathbb{N}}$.  I show that $(p^{i})_{i \in \mathbb{N}} \sim (r^{i})_{i \in \mathbb{N}}$.  Choose $k \in \mathbb{N}_{0}$, then there exists an $i_{0}$ such that $i \geq i_{0}$ implies that $p^{k} \ll q^{i}$.  Also there exists an $i_{1}$ such that $i \geq i_{1}$ implies that $q^{i_{0}} \ll r^{i}$, hence $p^{k} \ll r^{i}$ for all $i \geq i_{1}$.  Similary, one can find an $\hat{i}_{1}$ such that $i \geq \hat{i}_{1}$ implies that $r^{k} \ll p^{i}$.  Taking the maximum of $i_{1}$ and $\hat{i}_{1}$ proves the claim.  The case in which all sequences are past timelike Cauchy is identical.  We are left to prove the case where one of them is from a different type than the other two.  I only prove the case where $(p^{i})_{i \in \mathbb{N}}$ and $(q^{i})_{i \in \mathbb{N}}$ are future timelike Cauchy and $(r^{i})_{i \in \mathbb{N}}$ is past timelike Cauchy, the other case being analogous.  I show first that $p^{i} \ll r^{j}$ for all $i,j >0$.  Suppose there exist $k, l > 0$ such that $p^{k} \notin I^{-} (r^{l})$ and let $i_{0}$ be such that $i \geq i_{0}$ implies that $p^{k} \ll q^{i}$.  But then $q^{i} \notin I^{-} (r^{l})$, which is a contradiction.  Remark that for all $z \in \mathcal{M}$, $p^{k} \ll z$ for all $k$ iff $q^{l} \ll z$ for all $l$.    This implies that it is impossible for $z^{1},z^{2}$ to exist such that $p^{k} \ll z^{1} \ll z^{2}$ for all $k$ and $z^{2} \notin \bigcup_{j \in \mathbb{N}} I^{+} (r^{j})$.  Moreover, $\bigcup_{j \in \mathbb{N}} I^{-}(p^{j}) = \bigcup_{j \in \mathbb{N}} I^{-}(q^{j})$, which implies it is impossible for $z^{1},z^{2}$ to exist such that $r^{k} \gg z^{1} \gg z^{2}$ for all $k$ and $z^{2} \notin \bigcup_{j \in \mathbb{N}} I^{-} (p^{j})$.  $\square$
\section{Appendix B}
I show that the $\mathcal{C}^{+}_{\alpha}$ property is stable.  Let $\epsilon > 0$, I show that $$\alpha (\epsilon) \leq \min_{p \in \mathcal{M}^{\textrm{str} \downarrow \epsilon}} \max_{r \in \overline{B_{D}(p , \epsilon )}} d(p,r).$$
Choose $(p_{i})_{i \in \mathbb{N}} \in \mathcal{M}^{\textrm{str} \downarrow \epsilon}$, $\alpha ( \epsilon ) > \delta  >0$ and $\delta > 4 \gamma > 0$ such that $\left| x - \epsilon \right| < \gamma$ implies that $\left| \alpha (x) - \alpha(\epsilon) \right| < \frac{\delta}{2}$.  Let $i$ be sufficiently large such that $\frac{1}{2^{i-3}} < \gamma$, $\left| D_{\mathcal{M}_{i}} (p_{i} , \partial_{\textrm{F}} \mathcal{M}_{i}) - \epsilon \right| < \frac{\gamma}{2}$ and $D((p_{j})_{j \in \mathbb{N}} , p_{i}) < \frac{3}{2^{i+1}}$, where $D$ denotes also the metric on the disjoint union $\bigcup_{i \in \mathbb{N}} \mathcal{M}_{i}$, constructed in the proof of theorem 6.  Then, there exists an $r_{i} \in B_{D_{\mathcal{M}_{i}}} (p_{i} , \epsilon - \frac{\gamma}{2})$ such that $d_{g_{i}} ( p_{i} ,r_{i}) > \alpha ( \epsilon ) - \frac{\delta}{2}$.   The final remarks of the same proof show that there exists a point $r \in \mathcal{M}^{\textrm{str}}$ such that $D(r,r_{i}) < \frac{5}{2^{i+1}}$.  Hence, $$D((p_{i})_{i \in \mathbb{N}} ,r ) < \epsilon - \frac{\gamma}{2} + \frac{3}{2^{i+1}} + \frac{5}{2^{i+1}} < \epsilon. $$   
Moreover, (\ref{imp3}) on page 18 implies that
$$ \left| d((p_{j})_{j \in \mathbb{N}} ,r ) - d_{g_{i}}(p_{i} , r_{i}) \right| < \frac{3}{2^{i-2}} + \frac{3}{2^{i+1}} + \frac{5}{2^{i+1}} = \frac{1}{2^{i-4}} < \frac{\delta}{2}. $$
Hence, $d((p_{j})_{j \in \mathbb{N}} ,r ) > \alpha ( \epsilon ) - \delta$.  This shows that for any such $\delta > 0$, we can find an $r(\delta) \in B_{D}((p_{i})_{i \in \mathbb{N}} , \epsilon)$ such that $d((p_{j})_{j \in \mathbb{N}} ,r ) > \alpha ( \epsilon ) - \delta$.  The compactness of the closed $\epsilon$-balls and the continuity of $d$ in the strong topology finish the proof.  
\section{Appendix C}
I prove the claim made at the end of theorem 4.  Choose $\mathcal{C}$ to be a countable dense subset of $\mathcal{M}_{1}$, then by the usual diagonalization argument we can find a subsequence $(n_{k})_{k \in \mathbb{N}}$ such that $\psi_{n_{k}}(p) \stackrel{k \rightarrow \infty}{\rightarrow} \psi (p)$ for all $p \in \mathcal{C}$.  Clearly, $\psi$ preserves the strong as well as the Lorentz metric.  Suppose $\psi ( \mathcal{C} )$ is not dense in $\mathcal{M}_{2}$.  Then, choose a countable, dense subset $\mathcal{D} \subset \mathcal{M}_{2}$ which contains $\psi ( \mathcal{C} )$.  Taking a subsequence of $(n_{k})_{k \in \mathbb{N}}$ (which we denote in the same way) if necessary, we obtain that $\zeta_{n_{k}}(q) \stackrel{k \rightarrow \infty}{\rightarrow} \zeta(q)$ for all $q \in \mathcal{D}$.  Obviously, $\zeta \circ \psi$ equals the identity on $\mathcal{C}$.  Since $\psi ( \mathcal{C} )$ was supposed not to be dense in $\mathcal{M}_{2}$, we can find an $\epsilon >0$ and a point $q \in \mathcal{D}$ such that $D_{\mathcal{M}_{2}} (q ,\psi(p)) \geq \epsilon$ for all $p \in \mathcal{C}$, but this is impossible since that would imply that $D_{\mathcal{M}_{1}} (\zeta(q),r) \geq \epsilon$ for all $r \in \mathcal{C}$.  Hence, we choose $\mathcal{D} = \psi( \mathcal{C})$ and $\psi \circ \zeta$ equals the identity on $\psi ( \mathcal{C} )$.  I show now that $\psi$ has a unique continuous extension.  Let $r \in \mathcal{M}_{1} \setminus \mathcal{C}$ and choose a Cauchy sequence $(r_{n})_{n \in \mathbb{N}}$ converging to $r$, then $(\psi(r_{n}))_{n \in \mathbb{N}}$ is a Cauchy sequence converging to, say, $\psi(r)$.  It is left as an easy exercise to the reader to show that $\psi(r)$ is defined independently of the Cauchy sequence.  I show now that $\psi_{n_{k}} (r) \stackrel{k \rightarrow \infty }{\rightarrow} \psi(r)$.  Choose $\epsilon > 0$ and let $p \in \mathcal{C}$ such that $D_{\mathcal{M}_{1}} (p,r) < \frac{\epsilon}{3}$.  Choose $k$ large enough that $n_{k} > \frac{12}{\epsilon}$ and $\psi_{n_{k}} (p) \in B_{D_{\mathcal{M}_{2}}}( \psi (r )) ,\frac{\epsilon}{3})$.  Then, $D_{\mathcal{M}_{2}}( \psi_{n_{k}} (p) , \psi_{n_{k}} (r )) < \frac{2 \epsilon}{3}$ and the triangle inequality implies that $D_{\mathcal{M}_{2}}( \psi(r) , \psi_{n_{k}}(r)) < \epsilon$ which finishes the proof.          
\section{Appendix D}
Let $(X,d)$ be a topological space where $d$ is a (pseudo)
distance and denote by $ \tau $ the corresponding locally compact
topology. It is an elementary fact that the open balls $B_{1/n } (
p )$ with radius $1/n : n \in \mathbb{N}_{0} $ around $p$ define a
countable basis for $ \tau $ in $p$.  In this appendix $I,J$ will
denote index sets.  A $(X, \tau)$ cover $C$ is defined as follows:
$$ C = \{ A_{i} | A_{i} \in \tau, i \in I \} $$
such that $$ \bigcup_{ i \in I } A_{i} = X .$$ If $C =  \{ A_{i} |
A_{i} \in \tau, i \in I \} , D = \{ B_{j} | B_{j} \in \tau, j \in
J \} $ are $(X, \tau)$ covers then we say that $C$ is finer than
or is a refinement of $D$, $ C < D$ if and only if $$ \forall i
\in I \quad \exists j \in J : A_{i} \subset B_{j} .$$  Next we
define a few operations on the set of covers $C(X, \tau )$:
\\
\textbf{Operations on covers}
\begin{itemize}
\item Let $C,D$ be as before,  $$ C \wedge D  = \{ A_{i} \cap B_{j} | A_{i}, B_{j} \in \tau \quad
i \in I, j \in J \}$$ $C \wedge D$ is obviously a cover, moreover
the doublet $ C(X ,\tau),\wedge $ is a commutative semigroup.
\item For $ A \subset X$ the star of $A$ with respect to $C$ is
defined as follows:
$$ St(A,C) = \cup_{ A_{i} \in C: A \cap A_{i} \neq \emptyset}
A_{i} $$
\item The star of $C, C^{*}$ is then defined as:
$$ C^{*} = \{ St(A_{i},C) | A_{i} \in C \} $$
Remark that $ C < C^{*} < C^{**} \ldots $ and that if $I $ is
finite then there exists a $n \in \mathbb{N} $ such that after $n$
star operations $C$ has become the trivial cover.
\end{itemize}
Using the topological basis of open balls, we can define
elementary covers $C_{n} \quad n \in \mathbb{N}_{0} $ as follows:
$$ C_{n} = \{ B_{1/n } ( p ) | p \in X \} $$
These elementary covers now define a subset $U$ of $C(X, \tau )$ :
$$ U = \{ C \in C(X, \tau )| \exists C_{n} : C_{n} < C \} $$ The set $U$
satisfies the following obvious properties:
\begin{enumerate}
\item If $ C \in U $ and $ C < D $ then $ D \in U $
\item If $C,D \in U $ then $ C \wedge D \in U $
\item If $ C \in U $ then $ \exists D \in U : D^{*} < C $
\end{enumerate}
From now on we take the above properties as a \textbf{definition}
for a \textbf{uniformity}:
\begin{deffie}
Let $X$ be a set, a cover $C $ is defined as:
$$ C = \{ A_{i} | A_{i} \subset X, i \in I \} $$
such that $$ \bigcup_{ i \in I } A_{i} = X $$ A collection of
covers $U$ is called a \textbf{uniformity} for $X$ if and only if
\begin{enumerate}
\item If $ C \in U $ and $ C < D $ then $ D \in U $
\item If $C,D \in U $ then $ C \wedge D \in U $
\item If $ C \in U $ then $ \exists D \in U : D^{*} < C $
\end{enumerate}
where all definitions of $ < , \wedge $ and $^{*}$ are independent
of $ \tau $.
\end{deffie}
It has been proven that any uniformity can be generated by a
family of pseudodistances \cite{Page}.  This indicates a
uniformity defines a topology.  For our applications we need a
different ingredient.
\begin{deffie}
Let $I$ be a directed net, and suppose $B_{i}(x) \subset X$
satisfy the following properties:
\begin{enumerate}
\item $x \in B_{i} ( x ) \quad \forall x \in X, i \in I $
\item If $ i \leq j $ then $ B_{i} (x) \subset B_{j} (x) \quad
\forall x \in X $
\item $ \forall i \in I , \exists j \in I$  such that $ \forall y \in
B_{j}(x): x \in B_{i} (y)$
\item $ \forall i \in I, \exists j \in I$ such that if $ z \in
B_{j} (y), y \in B_{j}(x) $ then $ z \in B_{i} (x) $.
\end{enumerate}
then we call the family of all $B_{i}(x)$ \textbf{a uniform
neighborhood system}.
\end{deffie}
Now it has been proven that if $ \{ B_{i}(x) | x \in X, i \in I
\}$ is a uniform neighborhood system then the family of covers:
$$ C_{i} = \{ B_{i} (x) | x \in X \} $$
$i \in I$ is a basis for a uniformity on $X$.  On
the other hand every uniformity can be constructed from a uniform
neighborhood system.
\\
The topology $ \tau_{U} $ defined by a uniformity $ U $, the
\textbf{uniform topology}, is constructed as follows:
$$ O(x) \in \tau_{U} \Longleftrightarrow \exists C \in U : St(x,C)
\subset O(x) $$ so $ \{ St(x,C) | x \in X , C \in U \} $ defines a
basis for the topology. The topology is Hausdorff if and only if $
\bigcap_{O(x) \in \tau_{U}} O(x) = \{x\} $ but it is not difficult
to see that this is equivalent with: $$\bigcap_{i \in I } B_{i}(x)
= \{x\}
$$ where $ \{ B_{i} (x) | i \in I, x \in X \} $ is the uniform
neighborhood system which generates $U$.  We state a few facts about
\textbf{quotient uniformities}.
\textbf{Terminology}
\begin{itemize}
\item Let $(X,U)$ and $(Y,V)$ be uniform spaces, a map $f : X
\rightarrow Y $ is \textbf{uniformly continuous} if and only if
$$ \forall C \in V : f^{ -1} ( C ) \in U $$
where for $ C = \{ A_{i} | i \in I \} $ , $ f^{-1} (C) = \{ f^{-1}
(A_{i}) | i \in I \} $.
\item A uniformity $ \tilde{U} $ on $X$ is finer than $U$ if and
only if every cover in $U$ belongs to $ \tilde{U} $.
\item Let $ \pi : X \rightarrow \tilde{X} $ be a surjective map
and $(X,U)$ a uniform space, the \textbf{quotient uniformity} $
\tilde{U} $ on $ \tilde{X} $ is the finest uniformity which makes
$ \pi $ uniformly continuous.
\end{itemize}
Notice that the existence of a quotient uniformity is guaranteed
by the lemma of Zorn, the uniqueness is immediate.  The obvious
question now is if $ \tau_{ \tilde{U}} $ is equal to the quotient
topology of $ \tau_{U} $.  The answer is in general no, but under
some special circumstances it works.
\begin{deffie}
A uniform neighborhood system $ \{ B_{i} (x) | x \in X, i \in I
\}$ is \textbf{compatible} with an equivalence relation on $X$ if
and only if
$$ \forall i \in I, x' \sim x \text{ and } y \in B_{i}(x)
\quad \exists y' \sim y : y' \in B_{i} ( x' )
$$
\end{deffie}
As envisaged, compatibility implies that $ \tau_{ \tilde{U}} $ is
equal to the quotient topology of $ \tau_{U} $.
\\
\begin{theo}
If $U$ is generated by $ \{ B_{i}(x) | i \in I, x \in X \} $ which
is compatible with $ \sim $  which is for example defined by a
surjective map, then the quotient uniformity $\tilde{U} $ on $
\tilde{X} = X / \sim $ is generated by the uniform neighborhood
system defined by:
$$ \tilde{B}_{i} ( \tilde{x} ) = \{ \tilde{y} | \exists x \in
\tilde{x} \text{ and } y \in \tilde{y} : y \in B_{i} ( x)\} $$ $
\forall \tilde{x} \in \tilde{X}, i \in I$.  Moreover $ \tau_{
\tilde{U}} $ is equal to the quotient topology of $ \tau_{U} $ and
a basis of neighborhoods of $ \tilde{x} \in \tilde{X} $ is $ \{
\tilde{B}_{i} ( \tilde{x} ) | i \in I \} $
\end{theo}
As mentioned, every uniformity can be generated by a family of
pseudodistances. In the case that the uniformity is generated by a
countable uniform neighborhood system, the topology is defined by
one pseudodistance, which is a distance when the uniformity is
Hausdorff.  Suppose $ C_{n} = \{ B_{n} (x) | x \in X \} $, $ n \in
\mathbb{N} $ , is a countable basis for a uniformity $U$, then we
can find a subsequence $ (n_{k})_{k} $ such that:
$$ \forall k, \quad w \in B_{n_{k}} ( z),  z \in B_{n_{k}} (y),  y \in B_{n_{k}} (x
) \Rightarrow  w \in B_{n_{k-1}} (x)$$ Assume $C_{n}$ is such a
basis.
\\
\begin{theo}
Let $ C_{n} $ be a countable basis of $U$, then with
$$ \rho (x,y) = \inf_{ \{ n \geq 0, y \in B_{n} (x) \} }
2^{-n} $$ the function
$$ d(x,y) = \inf_{ K \in \mathbb{N},   x_{k}} \sum_{ k = 1}^{K} \frac{1}{2} (
\rho (x_{k-1} , x_{k} ) + \rho ( x_{k} , x_{k-1} ) ) $$ is a
pseudodistance which generates $U$.  $ \{ x_{0} , \ldots , x_{K}
\} $ with $ x_{0} = x, x_{K} = y$ is a path in $X$. If $U$ is
Hausdorff then $d$ is a distance.
\end{theo}
Note that the function $d$ depends on the choice of basis $C_{n}$
and is therefore not canonical.                       
\section{Acknowledgements}
I wish to thank Norbert Van den Bergh and Frans Cantrijn for the careful reading and many suggestions which made this paper more digestible.  I am also indebted to my colleague Benny Malengier who made the beautiful pictures for me.  Last but not least, I wish to express my gratitude towards Luca Bombelli and Rafael Sorkin for many discussions about this topic.

\end{document}